\documentclass[twocolumn,aps,prl,floatfix,superscriptaddress]{revtex4-1}
\usepackage{amsmath,amssymb}
\usepackage[colorinlistoftodos, color=green!40, prependcaption]{todonotes}
\usepackage{graphicx,float}
\usepackage{times,txfonts}
\usepackage{textcomp}
\usepackage{color}
\usepackage{multirow}
\usepackage{color}
\usepackage{soul}
\usepackage{natbib}
\usepackage{nicefrac}
\usepackage{multirow}
\usepackage{blindtext}
\usepackage{hyperref}
\usepackage{changes}
\usepackage[export]{adjustbox}
\usepackage{siunitx}
\definecolor{hugoColor}{RGB}{59,134,255}

\begin{document}

\title{Copper Damascene Process-Based High-Performance Thin Film Lithium Tantalate Modulators}

\author{Mengxin Lin}
\affiliation{Institute of Physics, Swiss Federal Institute of Technology, Lausanne (EPFL), CH-1015 Lausanne, Switzerland}
\affiliation{Institute of Electrical and Micro Engineering (IEM), EPFL, CH-1015 Lausanne, Switzerland}

\author{Zihan Li}
\affiliation{Institute of Physics, Swiss Federal Institute of Technology, Lausanne (EPFL), CH-1015 Lausanne, Switzerland}
\affiliation{Institute of Electrical and Micro Engineering (IEM), EPFL, CH-1015 Lausanne, Switzerland}

\author{Alexander Kotz}
\affiliation{Institute of Photonics and Quantum Electronics (IPQ), Karlsruhe Institute of Technology (KIT), 76131 Karlsruhe, Germany}

\author{Hugo Larocque}
\affiliation{Institute of Physics, Swiss Federal Institute of Technology, Lausanne (EPFL), CH-1015 Lausanne, Switzerland}
\affiliation{Institute of Electrical and Micro Engineering (IEM), EPFL, CH-1015 Lausanne, Switzerland}

\author{Johann Riemensberger}
\affiliation{Institute of Physics, Swiss Federal Institute of Technology, Lausanne (EPFL), CH-1015 Lausanne, Switzerland}
\affiliation{Institute of Electrical and Micro Engineering (IEM), EPFL, CH-1015 Lausanne, Switzerland}

\author{Christian Koos}
\affiliation{Institute of Photonics and Quantum Electronics (IPQ), Karlsruhe Institute of Technology (KIT), 76131 Karlsruhe, Germany}

\author{Tobias J. Kippenberg}
\email{tobias.kippenberg@epfl.ch}
\affiliation{Institute of Physics, Swiss Federal Institute of Technology, Lausanne (EPFL), CH-1015 Lausanne, Switzerland}
\affiliation{Institute of Electrical and Micro Engineering (IEM), EPFL, CH-1015 Lausanne, Switzerland}

\begin{abstract}
\end{abstract}

\maketitle

\begin{figure*}[t]
	\centering
	\includegraphics[width=\linewidth]{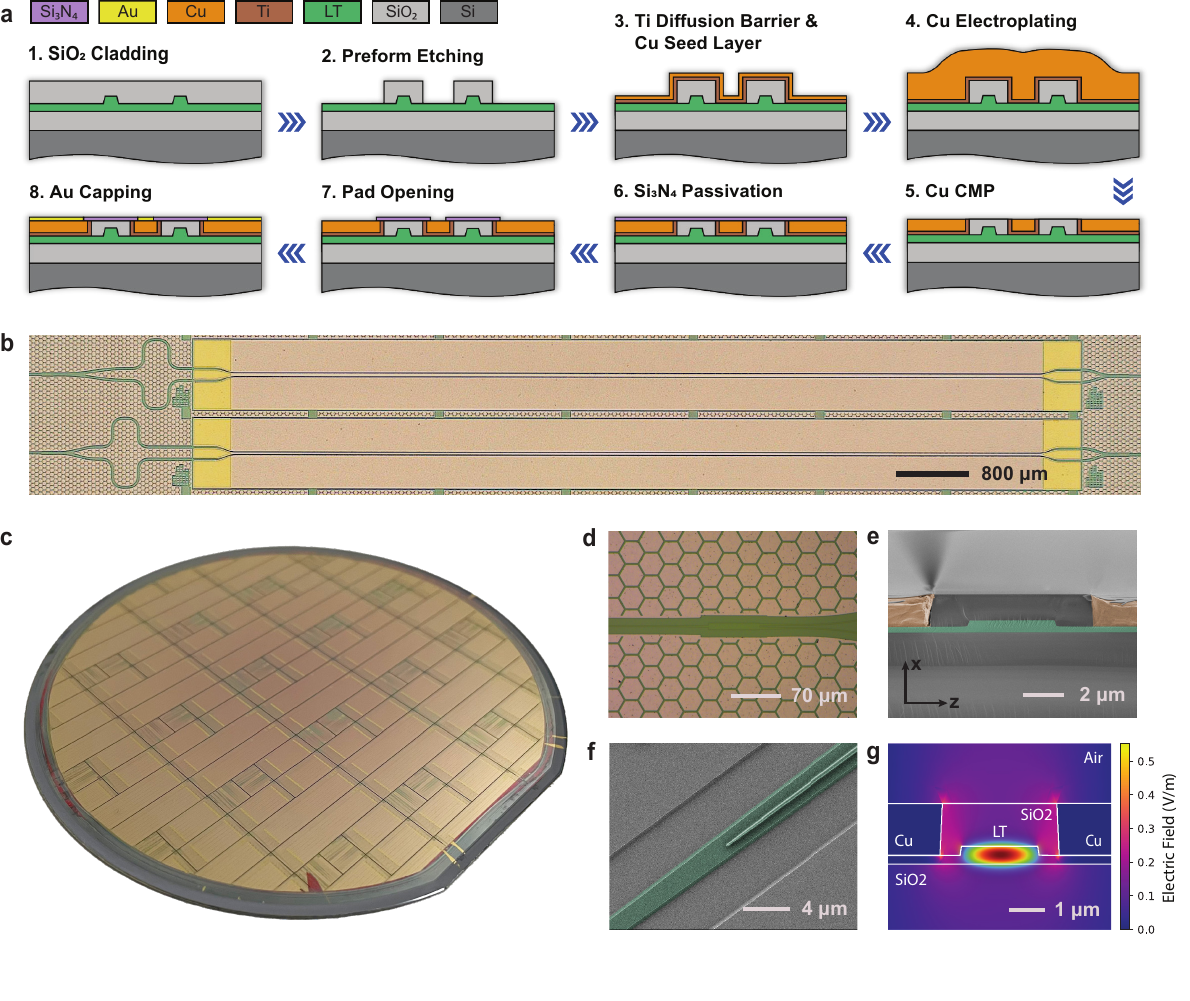}
	\caption{\textbf{Copper Damascene process-based thin film lithium tantalate (Cu-TFLT) Mach-Zehnder modulators (MZMs).} {\bf a,} Process flow for fabricating copper electrodes on patterned TFLT wafers using a Damascene process. {\bf b,} Microscope image of two Cu-TFLT MZMs, each comprising of two 1x2 multimode interference (MMI) couplers, an unbalanced arm, and a pair of push-pull phase shifters. The yellow color indicates the gold-capped pads for stable and efficient electrode probing. {\bf c,} Photograph of a manufactured 4-inch wafer hosting hundreds of Cu-TFLT MZMs. {\bf d,} Microscope image of an MMI coupler surrounded by honeycomb copper fillers, which are used to ensure uniform planarization. (e) Colored scanning electron microscopy (SEM) image of the cross-section of the lithium tantalate waveguide (green) and copper electrodes (orange). The waveguide has a rib height of \SI{320}{\nano\meter} and a slab thickness of \SI{280}{\nano\meter}. The copper electrode is \SI{1.8}{\micro\meter} thick and is passivated with \SI{100}{\nano\meter} of silicon nitride to avoid oxidation. {\bf f,} Colored SEM image of a double-layer taper for efficient and broadband edge coupling between the chip and single-mode fibers (coupling loss around 3 dB per facet). The smallest patterned feature has a dimension of \SI{200}{\nano\meter}. (g) Numerically simulated microwave and optical field distributions in the cross-section of the Cu-TFLT MZM.}
	\label{fig:fig1}
\end{figure*}

\textbf{Interfacing electrical and optical systems is a ubiquitous requirement for modern networks. Increasing data traffic and enabling greater connectivity between microelectronics are adding increasingly stringent requirements to devices capable of transferring information between these two types of systems. Competitive footprint, efficiency, and bandwidth figures have propelled interest in deploying integrated electro-optic modulators based on Pockels materials for such tasks. Due to its wide usage in legacy bulk electro-optic modulators, and triggered by the availability of 'on insulator' wafers, lithium niobate based devices have seen major advances. Recently, even more favorable properties have been demonstrated in lithium tantalate based devices, featuring similar Pockels effect, but exhibiting lower bias drift and lower birefringence, while equally benefiting from existing volume usage in wireless RF filters. 
Despite major progress in integrated modulators, these newly emerged ferro-electrical modulators cannot be integrated tightly with electronics yet using standardized processes, such as flip-chip bonding.  The latter is crucial to ensure low RF loss, and to obtain tight electronic-photonic integration. Indeed, such chip-on-chip or chip-on-wafer (CoW) hybrid bonding is already used for advanced GPU or CPU in microelectronics and heavily rely on standardized fabrication flows like the copper Damascene process.
Here, we overcome this bottleneck by incorporating the copper Damascene process in the fabrication of integrated lithium tantalate modulators. The copper Damascene process provides a flat surface in addition to low loss electrodes, enabling direct chip-on-wafer bonding for creating stacked chiplets. 
We demonstrate modulators featuring microwave losses that are $\sim\!10$\% lower than in designs relying on conventional gold electrodes. Our results allow us to reach data transmission figures on par with those of electro-optic modulators fabricated with other low-resistivity, yet less common, metals. Specifically, our fabricated modulators are able to achieve data transmission rates of 416 and 540 Gbit/s while preserving bit error ratios below the \SI{25}{\percent} SD-FEC threshold in a PAM4 and PAM8 transmission scheme, respectively. Together with the commercial availability of lithium tantalate as a Pockels material, our results open a path towards scalable and direct chip-on-wafer embedding of EO modulators with micro-electronic integrated circuits.}


\vspace{5mm}
Integrated electro-optic (EO) modulators offer a scalable platform for exchanging information between electrical and optical signals, thus providing alternative building blocks for high-capacity optical communications~\cite{Schmogrow:12,Pfeifle:14} while enabling new technologies such as chip-to-chip interconnects~\cite{Sun:15,Rizzo:23}. Advances in integrated EO modulators based on Pockels materials, such as lithium niobate and lithium tantalate~\cite{Li:23,WangNature:24}, have pushed the boundaries of this architecture in terms of device compactness, modulation efficiency, and bandwidth~\cite{Wang:18, He:19}. Further improvements on conventional Mach-Zehnder modulator (MZM) based architectures can involve modifications in their photonic or electrode design. For instance, resonator-based structures~\cite{Wang:18_2,Li:20,Larocque:24} can improve modulation efficiency and reduce footprint, though it comes at the expense of a voltage-bandwidth trade-off. Alternatively, incorporating micro-structured electrode designs, such as capacitively loaded coplanar waveguide (CPW) electrodes, can further decrease microwave losses and increase bandwidth~\cite{slowCPW:1994, Kharel:21}. Additional improvement in the bandwidth can rely on new electrode materials enabling lower propagation losses for the microwave signals interacting with the underlying optical waves. For example, manufacturing CPW electrodes with a low-resistivity material like silver can lead to modulators capable of data transmission rates in the hundreds of Gbit/s~\cite{WangOptica:24}.

Deploying such high-performance integrated Pockels EO modulators at scale will require volume manufacturing methods for photonic integrated circuits (PICs) that simultaneously feature high-quality photonic components and microwave electrodes. Though the photonic component of these Pockels PICs has reached this level of manufacturing~\cite{Luke:20, Li:23, WangNature:24}, their electrodes fall short compared to what commercial silicon PIC foundries can now provide~\cite{Fahrenkopf:19}. Specifically, implementing low-loss electrodes in Pockels PICs while ensuring compatibility with standardized fabrication flows adopted in industry remains an open challenge. Among such processes, copper Damascene patterning provides unique opportunities in semiconductor circuits by enabling greater multilayer connectivity between components such as transistors and electronic routing layers. Further process developments can also lead to the formation of tight-pitch copper interconnects between planarized dies and wafers through annealing-activated thermal expansion~\cite{Murugesan:21, Chew:23, Hahn:24}. Such techniques effectively overcome parasitic effects pervasive in bump-bonding and have proven instrumental in the assembly of hybrid-bonded systems such as chiplet-stacked GPUs and CPUs~\cite{Moore:24}. 

Here, we address this challenge by introducing the copper Damascene process in the fabrication of lithium tantalate PICs. Our process draws on mature, low-cost, and widespread methods producing electronic structures in microelectronics and silicon photonics~\cite{Fahrenkopf:19}. With this method, we demonstrate integrated EO modulators featuring $\sim\!10$\% lower microwave losses over devices relying on conventional gold electrodes, while maintaining the benefits of the Pockels effect in terms of modulation efficiency ($V_\pi L$). Equally important, the flat surface produced by the copper Damascene process potentially enables direct application of chip-on-chip or chip-on-wafer bonding techniques.
We then illustrate the practicality of such devices in high-rate pulse amplitude modulation (PAM) data transmission, where the performance of our devices reaches that of modulators relying on other low resistivity, yet less common, materials such as silver~\cite{WangOptica:24}.  



\vspace{5mm}
\noindent\textbf{Fabrication.} The devices were fabricated from a commercially available X-cut thin-film lithium tantalate (TFLT) wafer (NANOLN), which consists of \SI{600}{\nano\meter} of lithium tantalate ($\mathrm{LiTaO_3}$, LT), \SI{4.7}{\micro\meter} of $\mathrm{SiO_2}$, and a \SI{525}{\micro\meter}-thick high-resistivity silicon substrate. The TFLT PICs were manufactured using an etching technique based on a diamond-like carbon hard mask, which can reliably produce both lithium niobate and lithium tantalate PICs~\cite{Li:23, WangNature:24}. The etch depth of the LT was \SI{320}{\nano\meter}, leaving a \SI{280}{\nano\meter}-thick slab for efficient electro-optic modulation and proper phase matching between microwave and optical signals. Further details on the fabrication flow are provided in the Supplementary.

Next, the copper Damascene process, outlined in Fig.~\ref{fig:fig1}~(a), was introduced to fabricate low-loss traveling-wave electrodes. The process relies on the etching of electrode inlays in the silicon dioxide cladding followed by copper electroplating and chemical mechanical planarization (CMP). This process produces well defined electrodes with a smooth surface potentially suitable for chip-on-wafer bonding. Further information on the fabrication is available in the Methods. Figure~\ref{fig:fig1}~(b) shows two fabricated Cu-TFLT MZMs extending over an effective modulation length of \SI{6}{\milli\meter}. Figure~\ref{fig:fig1}~(c) illustrates the fabricated 4-inch wafer with hundreds of Cu-TFLT MZMs, highlighting its potential for scalable integration of next-generation photonic systems. To improve the wafer's uniformity following the CMP process, inactive regions feature a filler copper pattern, which is visible in Figs.~\ref{fig:fig1}~(b-d). As respectively shown in Figs.~\ref{fig:fig1}~(d-f), each modulator consists of 50:50 1$\times$2 multimode interference (MMI) couplers, EO phase shifters operating in a push-pull configuration, and double-layer tapers for enhanced PIC-to-fiber coupling efficiency~\cite{HeOL:19}. As illustrated in Fig.~\ref{fig:fig1}~(g), the tight confinement of the modulator's microwave and optical modes ensures strong overlap between the two fields and thus enhanced modulation efficiency. 

\begin{figure*}[t]
	\centering
	\includegraphics[width=\linewidth]{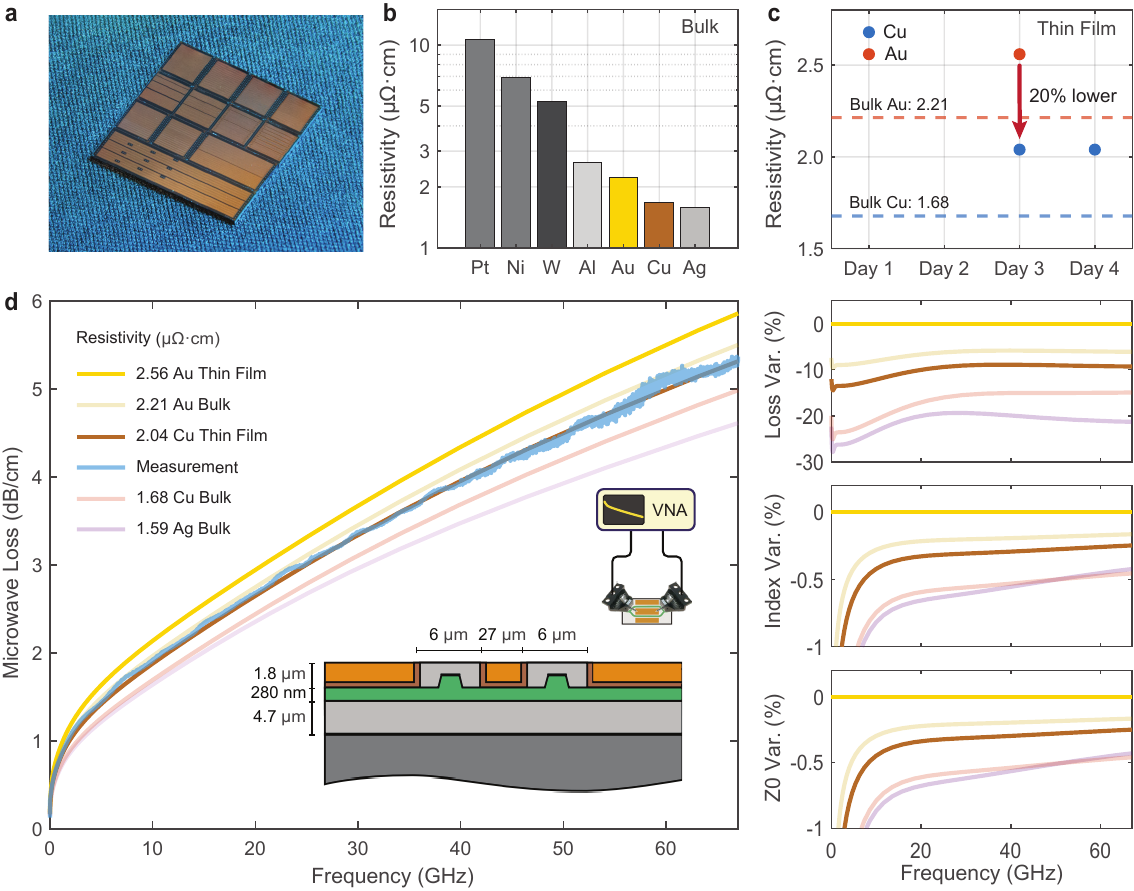}
	\caption{\textbf{Electrical characterization of Cu-TFLT MZMs} {\bf a,} A photograph of copper Damascene coplanar waveguide (CPW) electrode test structures fabricated on silicon. {\bf b,} Electrical resistivity of the 7 most conductive elemental metals, sorted by decreasing bulk room-temperature resistivity. {\bf c,} Measured resistivity of thin film gold and copper, and their evolution with time. The gold thin film (\SI{796}{\nano\meter}) is deposited by electron beam evaporation, and the copper thin film (\SI{1170}{\nano\meter}) is deposited by electroplating. Both metal thin films are deposited on silicon carrier wafers with \SI{2}{\micro\meter} of silicon dioxide, and measured using a combination of four-point probing and mechanical profilometry. The resistivity of the electroplated copper undergoes a \SI{24}{\percent} decrease within 48 hours at room temperature due to its self-annealing effect, and is stabilized at \SI{2.04}{\micro\ohm\cdot\centi\metre}, which is \SI{20}{\percent} lower than the measured resistivity of the gold thin film. {\bf d,} Measured microwave losses for the fabricated copper CPWs along with numerically simulated losses for electrodes composed of materials defined by different resistivities yet sharing the geometry of the fabricated device. Inset: microwave propagation loss testing apparatus and main dimensions used in the material stack of the measured device as previously defined in Fig.~\ref{fig:fig1}(a). Side panels: Corresponding percent loss, microwave index, and characteristic impedance ($Z_0$) variations of the simulated CPWs compared to thin film gold.}
	\label{fig:fig2}
\end{figure*}


\vspace{5mm}
\noindent\textbf{Electrical Transmission.} As outlined in the Supplementary, benchmarking copper Damascene CPW test structures manufactured on silicon wafers provided the ideal device parameters for the fabricated TFLT modulator. Figure~\ref{fig:fig2}~(a) provides a photograph of these test structures. These measurements confirmed improved microwave transmission figures of copper CPWs due to its lower resistivity. For reference, Fig.~\ref{fig:fig2}~(b) shows this resistivity and how it compares with those of other elemental metals commonly used in microelectronics and photonics. The resistivity values extracted from sheet resistance measurements shown in Fig.~\ref{fig:fig2}~(c) confirm the advantages of using copper over gold once it undergoes self-annealing effects over the course of 48 hours~\cite{Harper:99}. To gauge the performance of the fabricated CPWs, electrical characterization of its S-matrix parameters are used to derive properties such as microwave loss and effective index. In this work, the S-matrix was measured using a \SI{67}{\giga\hertz} vector network analyzer (Rohde \& Schwarz ZNA67). A pair of high-speed microwave probes (GGB), calibrated using the through-open-short-match (TOSM) standards, was used to launch microwave signals into the input port of the CPW and collect them from the output port. Figure~\ref{fig:fig2}~(d) shows the measured microwave loss of a \SI{16}{\milli\meter}-long CPW with a gap of \SI{6}{\micro\meter} and a center signal electrode width of \SI{27}{\micro\meter}. For reference, it also provides simulated data corresponding to CPWs with different resistivities sharing the geometric features of the fabricated device. This data suggests a 10\% microwave loss reduction achieved by using copper instead of gold. This lower resistivity also leads to a slightly lower microwave propagation index and characteristic electrode impedance.

\begin{figure*}[t]
	\centering
	\includegraphics[width=\linewidth]{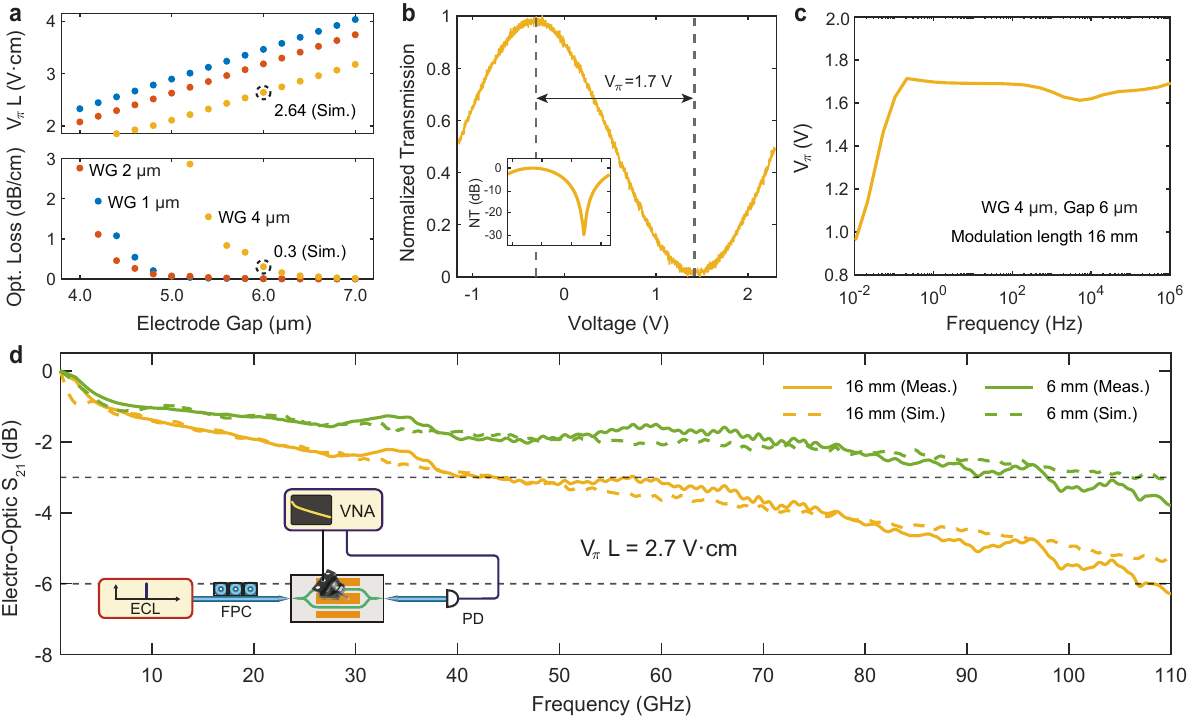}
         \caption{\textbf{Electro-optic characterization of Cu-TFLT MZMs.} {\bf a,} Numerical simulation of $V_\pi L$ and optical absorption loss of the MZMs for devices with different waveguide widths and electrode gap values. The simulated $V_\pi L$ is \SI{2.64}{\volt\cdot\centi\meter} for a device with a waveguide width of \SI{4}{\micro\meter} and an electrode gap of \SI{6}{\micro\meter}. {\bf b,} Normalized optical transmission of a \SI{16}{\milli\meter}-long device as a function of the applied voltage, showing a $V_\pi$ value of \SI{1.7}{\volt}. {\bf c,} Measured $V_\pi$ as a function of the sweeping frequency of the applied voltage signal. {\bf d,} Small-signal electro-optic response of devices with a modulation length of \SI{16}{\milli\meter} and \SI{6}{\milli\meter}, respectively. The measured $V_\pi L$ corresponds to the expected value in \textbf{a} for a \SI{4}{\micro\meter} waveguide width and a \SI{6}{\micro\meter} electrode gap. Inset: testing apparatus for the electro-optic response measurement.}
    \label{fig:fig3}
\end{figure*}

\vspace{5mm}
\noindent\textbf{Electro-optic Modulation.} A more straightforward way to evaluate the modulators' performance is to measure its electro-optic modulation properties, including the quasi-static half-wave voltage ($V_\pi$) and the 3~dB EO bandwidth. The quasi-static $V_\pi$ was measured by applying a low-frequency triangular voltage signal to the CPW electrode and recording the optical output simultaneously. Details regarding this procedure are provided in the Supplementary. 
To produce a device yielding optimal modulation figures, we perform finite element simulations (COMSOL Multiphysics) of the expected $V_\pi L$ and optical losses for various CPW geometries. Figure~\ref{fig:fig3}~(a) shows the results of these simulations, which determined the fabricated device geometry featuring a gap of \SI{6}{\micro\meter} and waveguide width of \SI{4}{\micro\meter}.
Figure~\ref{fig:fig3}~(b) shows the normalized transmission of a modulator as a function of the applied voltage (with a sweeping frequency of \SI{100}{\hertz}), highlighting an ideal sinusoidal shape and a $V_\pi$ of \SI{1.7}{\volt}. Given that the measured device has an effective modulation length of \SI{16}{\milli\meter}, this corresponds to a $V_\pi L$ of \SI{2.7}{\volt\cdot\centi\meter}, slightly higher than the simulated value of \SI{2.64}{\volt\cdot\centi\meter}. The inset dB-scale plot revealed an extinction ratio of \SI{30}{\decibel}, which is crucial for performing advanced communication experiments such as PAM4 and PAM8. The dependence of the $V_\pi$ on the frequency of the applied voltage signal was also studied, as shown in Figure~\ref{fig:fig3}~(c). Our modulators present a very stable $V_\pi$ across a broad range of frequencies from \SI{1}{\hertz} to \SI{1}{\mega\hertz}. This flat response is in contrast to that of previously reported lithium niobate MZMs with gold CPWs, which was observed to be very unstable below frequencies of \SI{100}{\kilo\hertz}~\cite{Holzgrafe:24}. With this stable $V_\pi$, heaters, which feature a stable response but high power consumption and crosstalk, are no longer needed for modulator biasing. Figure~\ref{fig:fig3}~(d) shows the measured EO response of our modulators up to \SI{110}{\giga\hertz}. The 3~dB EO bandwidth reaches \SI{40}{\giga\hertz} for a \SI{16}{\milli\meter}-long device, and is as high as \SI{100}{\giga\hertz} for a \SI{6}{\milli\meter} device, both featuring a CPW gap of \SI{6}{\micro\meter}. This superior EO modulation property indicates that our modulators can be used in high-speed communication systems to meet the continuously growing demands for high-speed data transmission. 

\begin{figure*}[t]
	\centering
	\includegraphics[width=\linewidth]{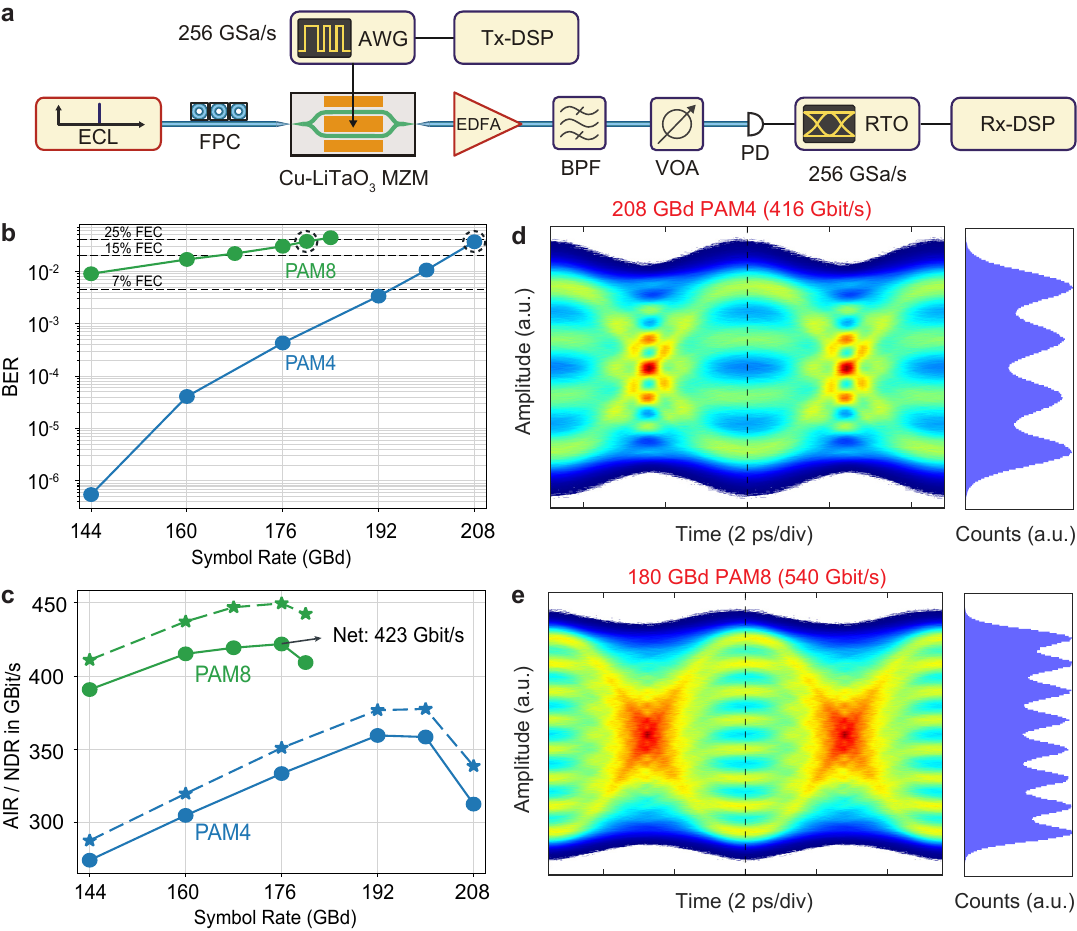}
	\caption{\textbf{Data transmission experiment using intensity-modulation direct detection (IMDD) with a Cu-TFLT MZM.} {\bf a,} Experimental apparatus: an external cavity laser (ECL) is used as the light source, and a fiber polarization controller (FPC) adjusts the polarization state of the light. Optical input and output coupling to the Cu-TFLT MZM is achieved using a pair of lensed fibers. The drive signals are synthesized by transmitter digital signal processing (Tx-DSP) and generated by an arbitrary waveform generator (AWG). The modulated optical signal is amplified using an erbium-doped fiber amplifier (EDFA), and out-of-band amplified spontaneous emission (ASE) noise is suppressed by a tunable bandpass filter (BPF). The amplified signal then passes through a variable optical attenuator (VOA) before being detected by a photodiode (PD). A high-speed real-time oscilloscope (RTO) samples the resulting signal, which is processed offline by receiver DSP (Rx-DSP). {\bf b,} Measured bit error ratios (BER) as a function of symbol rate for PAM8 (green) and PAM4 (blue) signals. The horizontal black dashed lines denote the thresholds for \SI{25}{\percent} and \SI{15}{\percent} soft-decision forward error correction (SD-FEC), as well as \SI{7}{\percent} hard-decision forward error correction (HD-FEC). {\bf c,} Extracted available information rates (AIR, dashed lines) and corresponding net data rates (NDR, solid lines) for measurements with BER values below the \SI{25}{\percent} SD-FEC threshold. The maximum NDR of 423 Gbit/s is obtained by using a PAM8 signal at a symbol rate of 176 GBd. {\bf d, e,} Eye diagrams and corresponding histograms, taken at the center of the symbol slot (indicated by the vertical dashed line), for the circled data points marked in {\bf b}.}	
    \label{fig:fig4}
\end{figure*}

\vspace{5mm}
\noindent\textbf{Optical Communications.}
To demonstrate the exceptional performance of our Cu-TFLT MZM, we performed a high-speed intensity-modulation and direct-detection (IMDD) signaling experiment. The experimental apparatus for this task is illustrated in Figure~\ref{fig:fig4}~(a). An external-cavity laser (ECL, \SI{17.8}{dBm} at \SI{1550}{\nano\meter}) provides the optical carrier. A high-speed arbitrary-waveform generator (AWG, M8199B, Keysight) is used to generate the electrical drive signal, which is fed to the CPW of the MZM via a \SI{20}{\centi\meter}-long RF cable, a broadband RF amplifier, and a \SI{110}{\giga\hertz} RF probe. We synthesize various pulse-amplitude modulation (PAM) signals based on pseudo-random bit sequences and apply root-raised cosine pulse-shaping filters with a roll-off of $\beta = 0.05$. We account for the frequency-dependent RF loss up to the input of the feeding probe by applying a linear minimum-mean-square-error (MMSE) predistortion. The CPW is terminated by a \SI{50}{\ohm} resistor via a second \SI{110}{\giga\hertz} RF probe. As previously demonstrated in Fig.~\ref{fig:fig3}(c), the EO stability of the Cu-TFLT modulator allows reliable DC biasing at the quadrature point for intensity modulation with a bias-T attached to the second probe. The optical power at the output fiber of the MZM operated at the quadrature point is \SI{5.6}{dBm}, which exceeds the power requirements in typical specifications for high-speed optical ethernet transceivers~\cite{ieee:17}. Still, an additional erbium-doped fiber amplifier (EDFA) was needed in the experiment to reach sufficient power levels for the high-speed photodiode (\SI{8.5}{dBm}) at the receiver. Note that a practical transceiver implementation could rely on a sufficiently broadband amplifier after the photodiode, thereby rendering the EDFA unnecessary. In our experiments, the out-of-band amplified spontaneous-emission (ASE) noise of the EDFA is suppressed by a tunable bandpass filter (BPF), and a variable optical attenuator (VOA) is used to adjust a power level of \SI{8.5}{dBm} at the input of the photodiode. The electrical signal at the photodiode output is digitized by a real-time oscilloscope (RTO, UXR 1004A, Keysight) with an analogue bandwidth of \SI{105}{\giga\hertz} and a sampling rate of 256 GSa/s. The data is finally extracted by offline receiver DSP (Rx-DSP), which contains resampling to 2 Sa/symbol, timing recovery, linear Sato equalization, and an additional decision-directed least-mean-square (DD-LMS) equalizer. In the data transmission experiment, we generated and received PAM4 and PAM8 data signals with symbol rates between \SI{144}{GBd} and \SI{208}{GBd}. Figure~\ref{fig:fig4}~(b) displays the measured bit error ratios (BER) of the various PAM signals as a function of symbol rate. The horizontal dashed lines indicate the thresholds for typical soft-decision (SD) forward-error correction (FEC) with \SI{25}{\percent} and \SI{15}{\percent} overhead, and for hard-decision (HD) FEC with \SI{7}{\percent} overhead. The results show that we can transmit \SI{180}{GBd} PAM8 signals and \SI{208}{GBd} PAM4 signals while the respective BER of \num{3.76e-2} and \num{3.56e-2} are still below the \SI{25}{\percent} SD-FEC limit. In Figure~\ref{fig:fig4}~(c), we further calculate the generalized mutual information (GMI) of our measurements based on log-likelihood ratios by using an additive white Gaussian noise channel model~\cite{Ivanov:16}. The dashed curves depict the achievable information rate (AIR), which is the product of the GMI and the symbol rate. The solid curves in Figure~\ref{fig:fig4}~(c) show the net data rates for measurements with BER below the \SI{25}{\percent} SD-FEC limit, where the net data rates are determined by multiplying the transmitted line rates with the code rate that is associated with the normalized GMI threshold, extracted from~\cite{Hu:22}. The results indicate that the highest AIR of \SI{449}{\giga\bit / \second} is achieved by using PAM8 signals at a symbol rate of \SI{176}{GBd}. This corresponds to a net data rate of \SI{423}{\giga\bit / \second}, which is on par with results demonstrated for high-bandwidth thin-film lithium niobate MZMs~\cite{Berikaa:23}. It should be noted that some of the results such as the PAM4 symbol rate were limited by the bandwidth of our driver electronics, not by the MZM itself.

\vspace{5mm}
\noindent\textbf{Discussion.} From a materials perspective, copper's low resistivity makes it ideally suited to minimize microwave losses in the manufactured modulators. As previously quantified in Fig.~\ref{fig:fig2}~(d), this lower resistivity improves microwave transmission compared to what can be achieved in materials like gold, which are more conventional yet feature a higher resistivity. Our results for electroplated copper also feature lower losses compared to copper-based modulators involving electron beam evaporation~\cite{Zhang:22}, provide a more practical approach to fabricating thicker metal structures, and finally introduce the possibility of fabricating multilayer electronics. This improvement is instrumental while using the modulator for data transmission. As shown in our pulse-amplitude modulation results from Fig.~\ref{fig:fig4}, modulators relying on copper CPWs can reach data transmission figures achieved with CPWs implemented with other low-resistivity metals like silver~\cite{WangOptica:24}. Further improvements in data transmission could involve incorporating our modulators within a transmitter architecture suitable for coherent communications~\cite{Xu:20, Xu:22}.

We can further reduce microwave losses with slight modifications of our PIC's material stack shown in Fig.~\ref{fig:fig1}(a). Our simulations suggest that some of these losses arise from the overlap between the propagating microwave field and our chip's silicon substrate. Implementing copper-based modulators on TFLT wafers with a larger bottom oxide layer separating the LT film from the substrate could therefore lead to further loss reduction. Finally, current crowding near the edge of the CPWs also contribute to our losses~\cite{Curran:10}. As further examined in the Supplementary, we can mitigate this effect by adopting a capacitively loaded CPW design, which can reduce microwave losses down to $\sim 1$~dB/cm at 50~GHz. Such electrodes do increase the effective index of the microwave field above that of the modulated optical field, thus resulting in a phase mismatch between the two fields that is not ideal for EO modulation. However, device modifications compatible with those that reduce other loss contributions, such as a thicker bottom oxide layer~\cite{Xue:24} or an undercut in the silicon substrate~\cite{Wang:22}, can reduce this mismatch and make low-loss capacitively loaded electrodes amenable to our copper Damascene platform. This improvement could lead to microwave losses closer to the 1~dB/cm figures measured in our capacitively loaded CPW test structures.

Besides lowering microwave losses, adopting a copper-based process for integrated EO modulators results in on-chip electronics inching closer to those used in commercially offered PICs manufactured in silicon foundries~\cite{Fahrenkopf:19}. Such similarities could prove useful in adapting emerging trends targeting three-dimensional (3D) silicon photonic integration based on copper bump bonding~\cite{Daudlin:25}. Here, the planarized surfaces resulting from the copper Damamascene process becomes instrumental for chip-on-wafer hybrid bonding, thereby providing the required connections in 3D-stacked chips. These techniques, already applied to the assembly of GPUs and CPUs, can now be employed with our lithium tantalate modulators.
Devices resulting from such dense electronics co-integration promise to find roles in fields ranging from data communications~\cite{Rizzo:23} to artificial intelligence~~\cite{Hua:25,Ahmed:25}. 

In summary, we introduced a copper Damascene process in the fabrication of TFLT EO modulators. Our process enables a 10\% reduction in microwave transmission loss compared to modulators relying on gold electrodes while maintaining a low $V_\pi$ and a broad bandwidth. Within data transmission settings, our copper-based modulators achieve the same metrics as those implemented with other less common low-resistivity metals. Their reliance on a metal-deposition process that is widely used in the semiconductor industry further reinforces the prospects of deploying co-packaged optics featuring next-generation EO PICs.

\section{Methods}

\noindent \textbf{Copper Damascene Fabrication Flow.} To begin with, the TFLT PICs were cladded with a \SI{2.5}{\micro\meter}-thick silicon dioxide layer by plasma-enhanced chemical vapor deposition (PECVD). This cladding layer will embed the subsequently formed copper electrodes. The layout for these electrodes was defined with a DUV stepper lithography (ASML PAS 5500/350C) and transferred as preforms into the cladding layer through fluorine-based dry etching in a reactive ion etcher. These preforms were coated with \SI{10}{\nano\meter} of titanium ($\mathrm{Ti}$) and \SI{100}{\nano\meter} of copper ($\mathrm{Cu}$) through sputtering (Alliance-Concept DP 650). The titanium layer serves as a barrier to prevent copper from diffusing into the surrounding dielectric~\cite{Edelstein:01}. It also provides adhesion to a subsequently sputtered copper layer acting as a seed to facilitate copper electroplating. The electroplating was performed in a Silicet Electroplating unit and contributed to a \SI{2.8}{\micro\meter} thick copper layer with low resistivity (\SI{2.04}{\micro\ohm\cdot\centi\metre}). Chemical mechanical planarization (CMP) was employed to remove the excess copper, leaving copper only in the etched preforms with a final thickness of \SI{1.8}{\micro\meter}. Subsequently, a \SI{100}{\nano\meter}-thick silicon nitride passivation layer was deposited with PECVD to prevent copper oxidation. The electrode pads were then opened and capped with gold (Au) for efficient and stable electrical probing. Finally, chip singulation was achieved through a combination of dry etching - standard reactive ion etching for LT and silicon dioxide, deep reactive ion etching for silicon - and backside grinding. This process ensures smooth facets for edge coupling to single-mode fibers.

\vspace{1 EM}

\noindent {\textbf{Data Availability}} {The data that support the plots within this paper and other findings of this study are available from the corresponding author upon request.}

\vspace{1 EM}

\noindent {\textbf{Code Availability}} {Code used to process the raw data into the results within this paper are available from the corresponding author upon request.}

\bibliography{main/CuLTOIBib}

\begin{thebibliography}{37}%
\makeatletter
\providecommand \@ifxundefined [1]{%
 \@ifx{#1\undefined}
}%
\providecommand \@ifnum [1]{%
 \ifnum #1\expandafter \@firstoftwo
 \else \expandafter \@secondoftwo
 \fi
}%
\providecommand \@ifx [1]{%
 \ifx #1\expandafter \@firstoftwo
 \else \expandafter \@secondoftwo
 \fi
}%
\providecommand \natexlab [1]{#1}%
\providecommand \enquote  [1]{``#1''}%
\providecommand \bibnamefont  [1]{#1}%
\providecommand \bibfnamefont [1]{#1}%
\providecommand \citenamefont [1]{#1}%
\providecommand \href@noop [0]{\@secondoftwo}%
\providecommand \href [0]{\begingroup \@sanitize@url \@href}%
\providecommand \@href[1]{\@@startlink{#1}\@@href}%
\providecommand \@@href[1]{\endgroup#1\@@endlink}%
\providecommand \@sanitize@url [0]{\catcode `\\12\catcode `\$12\catcode `\&12\catcode `\#12\catcode `\^12\catcode `\_12\catcode `\%12\relax}%
\providecommand \@@startlink[1]{}%
\providecommand \@@endlink[0]{}%
\providecommand \url  [0]{\begingroup\@sanitize@url \@url }%
\providecommand \@url [1]{\endgroup\@href {#1}{\urlprefix }}%
\providecommand \urlprefix  [0]{URL }%
\providecommand \Eprint [0]{\href }%
\providecommand \doibase [0]{http://dx.doi.org/}%
\providecommand \selectlanguage [0]{\@gobble}%
\providecommand \bibinfo  [0]{\@secondoftwo}%
\providecommand \bibfield  [0]{\@secondoftwo}%
\providecommand \translation [1]{[#1]}%
\providecommand \BibitemOpen [0]{}%
\providecommand \bibitemStop [0]{}%
\providecommand \bibitemNoStop [0]{.\EOS\space}%
\providecommand \EOS [0]{\spacefactor3000\relax}%
\providecommand \BibitemShut  [1]{\csname bibitem#1\endcsname}%
\let\auto@bib@innerbib\@empty
\bibitem [{\citenamefont {Schmogrow}\ \emph {et~al.}(2012)\citenamefont {Schmogrow}, \citenamefont {Hillerkuss}, \citenamefont {Wolf}, \citenamefont {B\"{a}uerle}, \citenamefont {Winter}, \citenamefont {Kleinow}, \citenamefont {Nebendahl}, \citenamefont {Dippon}, \citenamefont {Schindler}, \citenamefont {Koos}, \citenamefont {Freude},\ and\ \citenamefont {Leuthold}}]{Schmogrow:12}%
  \BibitemOpen
  \bibfield  {author} {\bibinfo {author} {\bibfnamefont {R.}~\bibnamefont {Schmogrow}}, \bibinfo {author} {\bibfnamefont {D.}~\bibnamefont {Hillerkuss}}, \bibinfo {author} {\bibfnamefont {S.}~\bibnamefont {Wolf}}, \bibinfo {author} {\bibfnamefont {B.}~\bibnamefont {B\"{a}uerle}}, \bibinfo {author} {\bibfnamefont {M.}~\bibnamefont {Winter}}, \bibinfo {author} {\bibfnamefont {P.}~\bibnamefont {Kleinow}}, \bibinfo {author} {\bibfnamefont {B.}~\bibnamefont {Nebendahl}}, \bibinfo {author} {\bibfnamefont {T.}~\bibnamefont {Dippon}}, \bibinfo {author} {\bibfnamefont {P.~C.}\ \bibnamefont {Schindler}}, \bibinfo {author} {\bibfnamefont {C.}~\bibnamefont {Koos}}, \bibinfo {author} {\bibfnamefont {W.}~\bibnamefont {Freude}}, \ and\ \bibinfo {author} {\bibfnamefont {J.}~\bibnamefont {Leuthold}},\ }\href {\doibase 10.1364/OE.20.006439} {\bibfield  {journal} {\bibinfo  {journal} {Opt. Express}\ }\textbf {\bibinfo {volume} {20}},\ \bibinfo {pages} {6439} (\bibinfo {year} {2012})}\BibitemShut {NoStop}%
\bibitem [{\citenamefont {Pfeifle}\ \emph {et~al.}(2014)\citenamefont {Pfeifle}, \citenamefont {Brasch}, \citenamefont {Lauermann}, \citenamefont {Yu}, \citenamefont {Wegner}, \citenamefont {Herr}, \citenamefont {Hartinger}, \citenamefont {Schindler}, \citenamefont {Li}, \citenamefont {Hillerkuss}, \citenamefont {Schmogrow}, \citenamefont {Weimann}, \citenamefont {Holzwarth}, \citenamefont {Freude}, \citenamefont {Leuthold}, \citenamefont {Kippenberg},\ and\ \citenamefont {Koos}}]{Pfeifle:14}%
  \BibitemOpen
  \bibfield  {author} {\bibinfo {author} {\bibfnamefont {J.}~\bibnamefont {Pfeifle}}, \bibinfo {author} {\bibfnamefont {V.}~\bibnamefont {Brasch}}, \bibinfo {author} {\bibfnamefont {M.}~\bibnamefont {Lauermann}}, \bibinfo {author} {\bibfnamefont {Y.}~\bibnamefont {Yu}}, \bibinfo {author} {\bibfnamefont {D.}~\bibnamefont {Wegner}}, \bibinfo {author} {\bibfnamefont {T.}~\bibnamefont {Herr}}, \bibinfo {author} {\bibfnamefont {K.}~\bibnamefont {Hartinger}}, \bibinfo {author} {\bibfnamefont {P.}~\bibnamefont {Schindler}}, \bibinfo {author} {\bibfnamefont {J.}~\bibnamefont {Li}}, \bibinfo {author} {\bibfnamefont {D.}~\bibnamefont {Hillerkuss}}, \bibinfo {author} {\bibfnamefont {R.}~\bibnamefont {Schmogrow}}, \bibinfo {author} {\bibfnamefont {C.}~\bibnamefont {Weimann}}, \bibinfo {author} {\bibfnamefont {R.}~\bibnamefont {Holzwarth}}, \bibinfo {author} {\bibfnamefont {W.}~\bibnamefont {Freude}}, \bibinfo {author} {\bibfnamefont {J.}~\bibnamefont {Leuthold}}, \bibinfo {author} {\bibfnamefont {T.~J.}\ \bibnamefont
  {Kippenberg}}, \ and\ \bibinfo {author} {\bibfnamefont {C.}~\bibnamefont {Koos}},\ }\href {\doibase 10.1038/nphoton.2014.57} {\bibfield  {journal} {\bibinfo  {journal} {Nat. Photonics}\ }\textbf {\bibinfo {volume} {8}},\ \bibinfo {pages} {375} (\bibinfo {year} {2014})}\BibitemShut {NoStop}%
\bibitem [{\citenamefont {Sun}\ \emph {et~al.}(2015)\citenamefont {Sun}, \citenamefont {Wade}, \citenamefont {Lee}, \citenamefont {Orcutt}, \citenamefont {Alloatti}, \citenamefont {Georgas}, \citenamefont {Waterman}, \citenamefont {Shainline}, \citenamefont {Avizienis}, \citenamefont {Lin}, \citenamefont {Moss}, \citenamefont {Kumar}, \citenamefont {Pavanello}, \citenamefont {Atabaki}, \citenamefont {Cook}, \citenamefont {Ou}, \citenamefont {Leu}, \citenamefont {Chen}, \citenamefont {Asanovi{\'c}}, \citenamefont {Ram}, \citenamefont {Popovi{\'c}},\ and\ \citenamefont {Stojanovi{\'c}}}]{Sun:15}%
  \BibitemOpen
  \bibfield  {author} {\bibinfo {author} {\bibfnamefont {C.}~\bibnamefont {Sun}}, \bibinfo {author} {\bibfnamefont {M.~T.}\ \bibnamefont {Wade}}, \bibinfo {author} {\bibfnamefont {Y.}~\bibnamefont {Lee}}, \bibinfo {author} {\bibfnamefont {J.~S.}\ \bibnamefont {Orcutt}}, \bibinfo {author} {\bibfnamefont {L.}~\bibnamefont {Alloatti}}, \bibinfo {author} {\bibfnamefont {M.~S.}\ \bibnamefont {Georgas}}, \bibinfo {author} {\bibfnamefont {A.~S.}\ \bibnamefont {Waterman}}, \bibinfo {author} {\bibfnamefont {J.~M.}\ \bibnamefont {Shainline}}, \bibinfo {author} {\bibfnamefont {R.~R.}\ \bibnamefont {Avizienis}}, \bibinfo {author} {\bibfnamefont {S.}~\bibnamefont {Lin}}, \bibinfo {author} {\bibfnamefont {B.~R.}\ \bibnamefont {Moss}}, \bibinfo {author} {\bibfnamefont {R.}~\bibnamefont {Kumar}}, \bibinfo {author} {\bibfnamefont {F.}~\bibnamefont {Pavanello}}, \bibinfo {author} {\bibfnamefont {A.~H.}\ \bibnamefont {Atabaki}}, \bibinfo {author} {\bibfnamefont {H.~M.}\ \bibnamefont {Cook}}, \bibinfo {author} {\bibfnamefont
  {A.~J.}\ \bibnamefont {Ou}}, \bibinfo {author} {\bibfnamefont {J.~C.}\ \bibnamefont {Leu}}, \bibinfo {author} {\bibfnamefont {Y.-H.}\ \bibnamefont {Chen}}, \bibinfo {author} {\bibfnamefont {K.}~\bibnamefont {Asanovi{\'c}}}, \bibinfo {author} {\bibfnamefont {R.~J.}\ \bibnamefont {Ram}}, \bibinfo {author} {\bibfnamefont {M.}~\bibnamefont {Popovi{\'c}}}, \ and\ \bibinfo {author} {\bibfnamefont {V.~M.}\ \bibnamefont {Stojanovi{\'c}}},\ }\href {\doibase 10.1038/nature16454} {\bibfield  {journal} {\bibinfo  {journal} {Nature}\ }\textbf {\bibinfo {volume} {528}},\ \bibinfo {pages} {534} (\bibinfo {year} {2015})}\BibitemShut {NoStop}%
\bibitem [{\citenamefont {Rizzo}\ \emph {et~al.}(2023)\citenamefont {Rizzo}, \citenamefont {Novick}, \citenamefont {Gopal}, \citenamefont {Kim}, \citenamefont {Ji}, \citenamefont {Daudlin}, \citenamefont {Okawachi}, \citenamefont {Cheng}, \citenamefont {Lipson}, \citenamefont {Gaeta},\ and\ \citenamefont {Bergman}}]{Rizzo:23}%
  \BibitemOpen
  \bibfield  {author} {\bibinfo {author} {\bibfnamefont {A.}~\bibnamefont {Rizzo}}, \bibinfo {author} {\bibfnamefont {A.}~\bibnamefont {Novick}}, \bibinfo {author} {\bibfnamefont {V.}~\bibnamefont {Gopal}}, \bibinfo {author} {\bibfnamefont {B.~Y.}\ \bibnamefont {Kim}}, \bibinfo {author} {\bibfnamefont {X.}~\bibnamefont {Ji}}, \bibinfo {author} {\bibfnamefont {S.}~\bibnamefont {Daudlin}}, \bibinfo {author} {\bibfnamefont {Y.}~\bibnamefont {Okawachi}}, \bibinfo {author} {\bibfnamefont {Q.}~\bibnamefont {Cheng}}, \bibinfo {author} {\bibfnamefont {M.}~\bibnamefont {Lipson}}, \bibinfo {author} {\bibfnamefont {A.~L.}\ \bibnamefont {Gaeta}}, \ and\ \bibinfo {author} {\bibfnamefont {K.}~\bibnamefont {Bergman}},\ }\href {\doibase 10.1038/s41566-023-01244-7} {\bibfield  {journal} {\bibinfo  {journal} {Nat. Photonics}\ }\textbf {\bibinfo {volume} {17}},\ \bibinfo {pages} {781} (\bibinfo {year} {2023})}\BibitemShut {NoStop}%
\bibitem [{\citenamefont {Li}\ \emph {et~al.}(2023)\citenamefont {Li}, \citenamefont {Wang}, \citenamefont {Lihachev}, \citenamefont {Zhang}, \citenamefont {Tan}, \citenamefont {Churaev}, \citenamefont {Kuznetsov}, \citenamefont {Siddharth}, \citenamefont {Bereyhi}, \citenamefont {Riemensberger},\ and\ \citenamefont {Kippenberg}}]{Li:23}%
  \BibitemOpen
  \bibfield  {author} {\bibinfo {author} {\bibfnamefont {Z.}~\bibnamefont {Li}}, \bibinfo {author} {\bibfnamefont {R.~N.}\ \bibnamefont {Wang}}, \bibinfo {author} {\bibfnamefont {G.}~\bibnamefont {Lihachev}}, \bibinfo {author} {\bibfnamefont {J.}~\bibnamefont {Zhang}}, \bibinfo {author} {\bibfnamefont {Z.}~\bibnamefont {Tan}}, \bibinfo {author} {\bibfnamefont {M.}~\bibnamefont {Churaev}}, \bibinfo {author} {\bibfnamefont {N.}~\bibnamefont {Kuznetsov}}, \bibinfo {author} {\bibfnamefont {A.}~\bibnamefont {Siddharth}}, \bibinfo {author} {\bibfnamefont {M.~J.}\ \bibnamefont {Bereyhi}}, \bibinfo {author} {\bibfnamefont {J.}~\bibnamefont {Riemensberger}}, \ and\ \bibinfo {author} {\bibfnamefont {T.~J.}\ \bibnamefont {Kippenberg}},\ }\href {\doibase 10.1038/s41467-023-40502-8} {\bibfield  {journal} {\bibinfo  {journal} {Nat. Commun.}\ }\textbf {\bibinfo {volume} {14}},\ \bibinfo {pages} {4856} (\bibinfo {year} {2023})}\BibitemShut {NoStop}%
\bibitem [{\citenamefont {Wang}\ \emph {et~al.}(2024{\natexlab{a}})\citenamefont {Wang}, \citenamefont {Li}, \citenamefont {Riemensberger}, \citenamefont {Lihachev}, \citenamefont {Churaev}, \citenamefont {Kao}, \citenamefont {Ji}, \citenamefont {Zhang}, \citenamefont {Blesin}, \citenamefont {Davydova}, \citenamefont {Chen}, \citenamefont {Huang}, \citenamefont {Wang}, \citenamefont {Ou},\ and\ \citenamefont {Kippenberg}}]{WangNature:24}%
  \BibitemOpen
  \bibfield  {author} {\bibinfo {author} {\bibfnamefont {C.}~\bibnamefont {Wang}}, \bibinfo {author} {\bibfnamefont {Z.}~\bibnamefont {Li}}, \bibinfo {author} {\bibfnamefont {J.}~\bibnamefont {Riemensberger}}, \bibinfo {author} {\bibfnamefont {G.}~\bibnamefont {Lihachev}}, \bibinfo {author} {\bibfnamefont {M.}~\bibnamefont {Churaev}}, \bibinfo {author} {\bibfnamefont {W.}~\bibnamefont {Kao}}, \bibinfo {author} {\bibfnamefont {X.}~\bibnamefont {Ji}}, \bibinfo {author} {\bibfnamefont {J.}~\bibnamefont {Zhang}}, \bibinfo {author} {\bibfnamefont {T.}~\bibnamefont {Blesin}}, \bibinfo {author} {\bibfnamefont {A.}~\bibnamefont {Davydova}}, \bibinfo {author} {\bibfnamefont {Y.}~\bibnamefont {Chen}}, \bibinfo {author} {\bibfnamefont {K.}~\bibnamefont {Huang}}, \bibinfo {author} {\bibfnamefont {X.}~\bibnamefont {Wang}}, \bibinfo {author} {\bibfnamefont {X.}~\bibnamefont {Ou}}, \ and\ \bibinfo {author} {\bibfnamefont {T.~J.}\ \bibnamefont {Kippenberg}},\ }\href {\doibase 10.1038/s41586-024-07369-1} {\bibfield  {journal}
  {\bibinfo  {journal} {Nature}\ }\textbf {\bibinfo {volume} {629}},\ \bibinfo {pages} {784} (\bibinfo {year} {2024}{\natexlab{a}})}\BibitemShut {NoStop}%
\bibitem [{\citenamefont {Wang}\ \emph {et~al.}(2018{\natexlab{a}})\citenamefont {Wang}, \citenamefont {Zhang}, \citenamefont {Chen}, \citenamefont {Bertrand}, \citenamefont {Shams-Ansari}, \citenamefont {Chandrasekhar}, \citenamefont {Winzer},\ and\ \citenamefont {Lončar}}]{Wang:18}%
  \BibitemOpen
  \bibfield  {author} {\bibinfo {author} {\bibfnamefont {C.}~\bibnamefont {Wang}}, \bibinfo {author} {\bibfnamefont {M.}~\bibnamefont {Zhang}}, \bibinfo {author} {\bibfnamefont {X.}~\bibnamefont {Chen}}, \bibinfo {author} {\bibfnamefont {M.}~\bibnamefont {Bertrand}}, \bibinfo {author} {\bibfnamefont {A.}~\bibnamefont {Shams-Ansari}}, \bibinfo {author} {\bibfnamefont {S.}~\bibnamefont {Chandrasekhar}}, \bibinfo {author} {\bibfnamefont {P.}~\bibnamefont {Winzer}}, \ and\ \bibinfo {author} {\bibfnamefont {M.}~\bibnamefont {Lončar}},\ }\href {\doibase 10.1038/s41586-018-0551-y} {\bibfield  {journal} {\bibinfo  {journal} {Nature}\ }\textbf {\bibinfo {volume} {562}},\ \bibinfo {pages} {101} (\bibinfo {year} {2018}{\natexlab{a}})}\BibitemShut {NoStop}%
\bibitem [{\citenamefont {He}\ \emph {et~al.}(2019{\natexlab{a}})\citenamefont {He}, \citenamefont {Xu}, \citenamefont {Ren}, \citenamefont {Jian}, \citenamefont {Ruan}, \citenamefont {Xu}, \citenamefont {Gao}, \citenamefont {Sun}, \citenamefont {Wen}, \citenamefont {Zhou}, \citenamefont {Liu}, \citenamefont {Guo}, \citenamefont {Chen}, \citenamefont {Yu}, \citenamefont {Liu},\ and\ \citenamefont {Cai}}]{He:19}%
  \BibitemOpen
  \bibfield  {author} {\bibinfo {author} {\bibfnamefont {M.}~\bibnamefont {He}}, \bibinfo {author} {\bibfnamefont {M.}~\bibnamefont {Xu}}, \bibinfo {author} {\bibfnamefont {Y.}~\bibnamefont {Ren}}, \bibinfo {author} {\bibfnamefont {J.}~\bibnamefont {Jian}}, \bibinfo {author} {\bibfnamefont {Z.}~\bibnamefont {Ruan}}, \bibinfo {author} {\bibfnamefont {Y.}~\bibnamefont {Xu}}, \bibinfo {author} {\bibfnamefont {S.}~\bibnamefont {Gao}}, \bibinfo {author} {\bibfnamefont {S.}~\bibnamefont {Sun}}, \bibinfo {author} {\bibfnamefont {X.}~\bibnamefont {Wen}}, \bibinfo {author} {\bibfnamefont {L.}~\bibnamefont {Zhou}}, \bibinfo {author} {\bibfnamefont {L.}~\bibnamefont {Liu}}, \bibinfo {author} {\bibfnamefont {C.}~\bibnamefont {Guo}}, \bibinfo {author} {\bibfnamefont {H.}~\bibnamefont {Chen}}, \bibinfo {author} {\bibfnamefont {S.}~\bibnamefont {Yu}}, \bibinfo {author} {\bibfnamefont {L.}~\bibnamefont {Liu}}, \ and\ \bibinfo {author} {\bibfnamefont {X.}~\bibnamefont {Cai}},\ }\href {\doibase 10.1038/s41566-019-0378-6}
  {\bibfield  {journal} {\bibinfo  {journal} {Nat. Photonics}\ }\textbf {\bibinfo {volume} {13}},\ \bibinfo {pages} {359} (\bibinfo {year} {2019}{\natexlab{a}})}\BibitemShut {NoStop}%
\bibitem [{\citenamefont {Wang}\ \emph {et~al.}(2018{\natexlab{b}})\citenamefont {Wang}, \citenamefont {Zhang}, \citenamefont {Stern}, \citenamefont {Lipson},\ and\ \citenamefont {Lon\v{c}ar}}]{Wang:18_2}%
  \BibitemOpen
  \bibfield  {author} {\bibinfo {author} {\bibfnamefont {C.}~\bibnamefont {Wang}}, \bibinfo {author} {\bibfnamefont {M.}~\bibnamefont {Zhang}}, \bibinfo {author} {\bibfnamefont {B.}~\bibnamefont {Stern}}, \bibinfo {author} {\bibfnamefont {M.}~\bibnamefont {Lipson}}, \ and\ \bibinfo {author} {\bibfnamefont {M.}~\bibnamefont {Lon\v{c}ar}},\ }\href {\doibase 10.1364/OE.26.001547} {\bibfield  {journal} {\bibinfo  {journal} {Opt. Express}\ }\textbf {\bibinfo {volume} {26}},\ \bibinfo {pages} {1547} (\bibinfo {year} {2018}{\natexlab{b}})}\BibitemShut {NoStop}%
\bibitem [{\citenamefont {Li}\ \emph {et~al.}(2020)\citenamefont {Li}, \citenamefont {Ling}, \citenamefont {He}, \citenamefont {Javid}, \citenamefont {Xue},\ and\ \citenamefont {Lin}}]{Li:20}%
  \BibitemOpen
  \bibfield  {author} {\bibinfo {author} {\bibfnamefont {M.}~\bibnamefont {Li}}, \bibinfo {author} {\bibfnamefont {J.}~\bibnamefont {Ling}}, \bibinfo {author} {\bibfnamefont {Y.}~\bibnamefont {He}}, \bibinfo {author} {\bibfnamefont {U.~A.}\ \bibnamefont {Javid}}, \bibinfo {author} {\bibfnamefont {S.}~\bibnamefont {Xue}}, \ and\ \bibinfo {author} {\bibfnamefont {Q.}~\bibnamefont {Lin}},\ }\href {\doibase 10.1038/s41467-020-17950-7} {\bibfield  {journal} {\bibinfo  {journal} {Nat. Commun.}\ }\textbf {\bibinfo {volume} {11}},\ \bibinfo {pages} {4123} (\bibinfo {year} {2020})}\BibitemShut {NoStop}%
\bibitem [{\citenamefont {Larocque}\ \emph {et~al.}(2024)\citenamefont {Larocque}, \citenamefont {Sludds}, \citenamefont {Sattari}, \citenamefont {Christen}, \citenamefont {Choong}, \citenamefont {Prieto}, \citenamefont {Leo}, \citenamefont {Zarebidaki}, \citenamefont {Lohani}, \citenamefont {Kirby}, \citenamefont {Soykal}, \citenamefont {Soltani}, \citenamefont {Ghadimi}, \citenamefont {Englund},\ and\ \citenamefont {Heuck}}]{Larocque:24}%
  \BibitemOpen
  \bibfield  {author} {\bibinfo {author} {\bibfnamefont {D.~L.~P.}\ \bibnamefont {Larocque}, \bibfnamefont {Hugoand~Vitullo}}, \bibinfo {author} {\bibfnamefont {A.}~\bibnamefont {Sludds}}, \bibinfo {author} {\bibfnamefont {H.}~\bibnamefont {Sattari}}, \bibinfo {author} {\bibfnamefont {I.}~\bibnamefont {Christen}}, \bibinfo {author} {\bibfnamefont {G.}~\bibnamefont {Choong}}, \bibinfo {author} {\bibfnamefont {I.}~\bibnamefont {Prieto}}, \bibinfo {author} {\bibfnamefont {J.}~\bibnamefont {Leo}}, \bibinfo {author} {\bibfnamefont {H.}~\bibnamefont {Zarebidaki}}, \bibinfo {author} {\bibfnamefont {S.}~\bibnamefont {Lohani}}, \bibinfo {author} {\bibfnamefont {B.~T.}\ \bibnamefont {Kirby}}, \bibinfo {author} {\bibfnamefont {{\"O}.}~\bibnamefont {Soykal}}, \bibinfo {author} {\bibfnamefont {M.}~\bibnamefont {Soltani}}, \bibinfo {author} {\bibfnamefont {A.~H.}\ \bibnamefont {Ghadimi}}, \bibinfo {author} {\bibfnamefont {D.~R.}\ \bibnamefont {Englund}}, \ and\ \bibinfo {author} {\bibfnamefont {M.}~\bibnamefont {Heuck}},\
  }\href@noop {} {\bibfield  {journal} {\bibinfo  {journal} {ACS Photonics}\ }\textbf {\bibinfo {volume} {11}},\ \bibinfo {pages} {3860} (\bibinfo {year} {2024})}\BibitemShut {NoStop}%
\bibitem [{\citenamefont {Spickermann}\ and\ \citenamefont {Dagli}(1994)}]{slowCPW:1994}%
  \BibitemOpen
  \bibfield  {author} {\bibinfo {author} {\bibfnamefont {R.}~\bibnamefont {Spickermann}}\ and\ \bibinfo {author} {\bibfnamefont {N.}~\bibnamefont {Dagli}},\ }\href {\doibase 10.1109/22.320774} {\bibfield  {journal} {\bibinfo  {journal} {IEEE Trans. Microw. Theory Techn.}\ }\textbf {\bibinfo {volume} {42}},\ \bibinfo {pages} {1918} (\bibinfo {year} {1994})}\BibitemShut {NoStop}%
\bibitem [{\citenamefont {Kharel}\ \emph {et~al.}(2021)\citenamefont {Kharel}, \citenamefont {Reimer}, \citenamefont {Luke}, \citenamefont {He},\ and\ \citenamefont {Zhang}}]{Kharel:21}%
  \BibitemOpen
  \bibfield  {author} {\bibinfo {author} {\bibfnamefont {P.}~\bibnamefont {Kharel}}, \bibinfo {author} {\bibfnamefont {C.}~\bibnamefont {Reimer}}, \bibinfo {author} {\bibfnamefont {K.}~\bibnamefont {Luke}}, \bibinfo {author} {\bibfnamefont {L.}~\bibnamefont {He}}, \ and\ \bibinfo {author} {\bibfnamefont {M.}~\bibnamefont {Zhang}},\ }\href {\doibase 10.1364/OPTICA.416155} {\bibfield  {journal} {\bibinfo  {journal} {Optica}\ }\textbf {\bibinfo {volume} {8}},\ \bibinfo {pages} {357} (\bibinfo {year} {2021})}\BibitemShut {NoStop}%
\bibitem [{\citenamefont {Wang}\ \emph {et~al.}(2024{\natexlab{b}})\citenamefont {Wang}, \citenamefont {Fang}, \citenamefont {Zhang}, \citenamefont {Kotz}, \citenamefont {Lihachev}, \citenamefont {Churaev}, \citenamefont {Li}, \citenamefont {Schwarzenberger}, \citenamefont {Ou}, \citenamefont {Koos},\ and\ \citenamefont {Kippenberg}}]{WangOptica:24}%
  \BibitemOpen
  \bibfield  {author} {\bibinfo {author} {\bibfnamefont {C.}~\bibnamefont {Wang}}, \bibinfo {author} {\bibfnamefont {D.}~\bibnamefont {Fang}}, \bibinfo {author} {\bibfnamefont {J.}~\bibnamefont {Zhang}}, \bibinfo {author} {\bibfnamefont {A.}~\bibnamefont {Kotz}}, \bibinfo {author} {\bibfnamefont {G.}~\bibnamefont {Lihachev}}, \bibinfo {author} {\bibfnamefont {M.}~\bibnamefont {Churaev}}, \bibinfo {author} {\bibfnamefont {Z.}~\bibnamefont {Li}}, \bibinfo {author} {\bibfnamefont {A.}~\bibnamefont {Schwarzenberger}}, \bibinfo {author} {\bibfnamefont {X.}~\bibnamefont {Ou}}, \bibinfo {author} {\bibfnamefont {C.}~\bibnamefont {Koos}}, \ and\ \bibinfo {author} {\bibfnamefont {T.~J.}\ \bibnamefont {Kippenberg}},\ }\href {\doibase 10.1364/OPTICA.537730} {\bibfield  {journal} {\bibinfo  {journal} {Optica}\ }\textbf {\bibinfo {volume} {11}},\ \bibinfo {pages} {1614} (\bibinfo {year} {2024}{\natexlab{b}})}\BibitemShut {NoStop}%
\bibitem [{\citenamefont {Luke}\ \emph {et~al.}(2020)\citenamefont {Luke}, \citenamefont {Kharel}, \citenamefont {Reimer}, \citenamefont {He}, \citenamefont {Loncar},\ and\ \citenamefont {Zhang}}]{Luke:20}%
  \BibitemOpen
  \bibfield  {author} {\bibinfo {author} {\bibfnamefont {K.}~\bibnamefont {Luke}}, \bibinfo {author} {\bibfnamefont {P.}~\bibnamefont {Kharel}}, \bibinfo {author} {\bibfnamefont {C.}~\bibnamefont {Reimer}}, \bibinfo {author} {\bibfnamefont {L.}~\bibnamefont {He}}, \bibinfo {author} {\bibfnamefont {M.}~\bibnamefont {Loncar}}, \ and\ \bibinfo {author} {\bibfnamefont {M.}~\bibnamefont {Zhang}},\ }\href {\doibase 10.1364/OE.401959} {\bibfield  {journal} {\bibinfo  {journal} {Opt. Express}\ }\textbf {\bibinfo {volume} {28}},\ \bibinfo {pages} {24452} (\bibinfo {year} {2020})}\BibitemShut {NoStop}%
\bibitem [{\citenamefont {Fahrenkopf}\ \emph {et~al.}(2019)\citenamefont {Fahrenkopf}, \citenamefont {McDonough}, \citenamefont {Leake}, \citenamefont {Su}, \citenamefont {Timurdogan},\ and\ \citenamefont {Coolbaugh}}]{Fahrenkopf:19}%
  \BibitemOpen
  \bibfield  {author} {\bibinfo {author} {\bibfnamefont {N.~M.}\ \bibnamefont {Fahrenkopf}}, \bibinfo {author} {\bibfnamefont {C.}~\bibnamefont {McDonough}}, \bibinfo {author} {\bibfnamefont {G.~L.}\ \bibnamefont {Leake}}, \bibinfo {author} {\bibfnamefont {Z.}~\bibnamefont {Su}}, \bibinfo {author} {\bibfnamefont {E.}~\bibnamefont {Timurdogan}}, \ and\ \bibinfo {author} {\bibfnamefont {D.~D.}\ \bibnamefont {Coolbaugh}},\ }\href {\doibase 10.1109/JSTQE.2019.2935698} {\bibfield  {journal} {\bibinfo  {journal} {IEEE J. Sel. Top. Quantum Electron.}\ }\textbf {\bibinfo {volume} {25}},\ \bibinfo {pages} {1} (\bibinfo {year} {2019})}\BibitemShut {NoStop}%
\bibitem [{\citenamefont {Murugesan}\ \emph {et~al.}(2021)\citenamefont {Murugesan}, \citenamefont {Sone}, \citenamefont {Simomura}, \citenamefont {Motoyoshi}, \citenamefont {Sawa}, \citenamefont {Fukuda}, \citenamefont {Koyanagi},\ and\ \citenamefont {Fukushima}}]{Murugesan:21}%
  \BibitemOpen
  \bibfield  {author} {\bibinfo {author} {\bibfnamefont {M.}~\bibnamefont {Murugesan}}, \bibinfo {author} {\bibfnamefont {E.}~\bibnamefont {Sone}}, \bibinfo {author} {\bibfnamefont {A.}~\bibnamefont {Simomura}}, \bibinfo {author} {\bibfnamefont {M.}~\bibnamefont {Motoyoshi}}, \bibinfo {author} {\bibfnamefont {M.}~\bibnamefont {Sawa}}, \bibinfo {author} {\bibfnamefont {K.}~\bibnamefont {Fukuda}}, \bibinfo {author} {\bibfnamefont {M.}~\bibnamefont {Koyanagi}}, \ and\ \bibinfo {author} {\bibfnamefont {T.}~\bibnamefont {Fukushima}},\ }in\ \href {\doibase 10.1109/3DIC52383.2021.9687604} {\emph {\bibinfo {booktitle} {2021 IEEE International 3D Systems Integration Conference (3DIC)}}}\ (\bibinfo {year} {2021})\ pp.\ \bibinfo {pages} {1--4}\BibitemShut {NoStop}%
\bibitem [{\citenamefont {Chew}\ \emph {et~al.}(2023)\citenamefont {Chew}, \citenamefont {Zhang}, \citenamefont {Vanstreels}, \citenamefont {Chery}, \citenamefont {De~Messemaeker}, \citenamefont {Witters}, \citenamefont {Van~Sever}, \citenamefont {Iacovo}, \citenamefont {Dewilde}, \citenamefont {Stucchi}, \citenamefont {De~Vos}, \citenamefont {Beyer}, \citenamefont {Miller},\ and\ \citenamefont {Beyne}}]{Chew:23}%
  \BibitemOpen
  \bibfield  {author} {\bibinfo {author} {\bibfnamefont {S.-A.}\ \bibnamefont {Chew}}, \bibinfo {author} {\bibfnamefont {B.}~\bibnamefont {Zhang}}, \bibinfo {author} {\bibfnamefont {K.}~\bibnamefont {Vanstreels}}, \bibinfo {author} {\bibfnamefont {E.}~\bibnamefont {Chery}}, \bibinfo {author} {\bibfnamefont {J.}~\bibnamefont {De~Messemaeker}}, \bibinfo {author} {\bibfnamefont {L.}~\bibnamefont {Witters}}, \bibinfo {author} {\bibfnamefont {K.}~\bibnamefont {Van~Sever}}, \bibinfo {author} {\bibfnamefont {S.}~\bibnamefont {Iacovo}}, \bibinfo {author} {\bibfnamefont {S.}~\bibnamefont {Dewilde}}, \bibinfo {author} {\bibfnamefont {M.}~\bibnamefont {Stucchi}}, \bibinfo {author} {\bibfnamefont {J.}~\bibnamefont {De~Vos}}, \bibinfo {author} {\bibfnamefont {G.}~\bibnamefont {Beyer}}, \bibinfo {author} {\bibfnamefont {A.}~\bibnamefont {Miller}}, \ and\ \bibinfo {author} {\bibfnamefont {E.}~\bibnamefont {Beyne}},\ }in\ \href {\doibase 10.1109/IEDM45741.2023.10413829} {\emph {\bibinfo {booktitle} {2023 International
  Electron Devices Meeting (IEDM)}}}\ (\bibinfo {year} {2023})\ pp.\ \bibinfo {pages} {1--4}\BibitemShut {NoStop}%
\bibitem [{\citenamefont {Hahn}\ \emph {et~al.}(2024)\citenamefont {Hahn}, \citenamefont {Kim}, \citenamefont {Lim}, \citenamefont {Moon},\ and\ \citenamefont {Rhee}}]{Hahn:24}%
  \BibitemOpen
  \bibfield  {author} {\bibinfo {author} {\bibfnamefont {S.~H.}\ \bibnamefont {Hahn}}, \bibinfo {author} {\bibfnamefont {W.}~\bibnamefont {Kim}}, \bibinfo {author} {\bibfnamefont {K.}~\bibnamefont {Lim}}, \bibinfo {author} {\bibfnamefont {B.}~\bibnamefont {Moon}}, \ and\ \bibinfo {author} {\bibfnamefont {M.}~\bibnamefont {Rhee}},\ }in\ \href {\doibase 10.1109/ECTC51529.2024.00215} {\emph {\bibinfo {booktitle} {2024 IEEE 74th Electronic Components and Technology Conference (ECTC)}}}\ (\bibinfo {year} {2024})\ pp.\ \bibinfo {pages} {1322--1328}\BibitemShut {NoStop}%
\bibitem [{\citenamefont {Moore}(2024)}]{Moore:24}%
  \BibitemOpen
  \bibfield  {author} {\bibinfo {author} {\bibfnamefont {S.~K.}\ \bibnamefont {Moore}},\ }\href {\doibase 10.1109/MSPEC.2024.10669241} {\bibfield  {journal} {\bibinfo  {journal} {IEEE Spectrum}\ }\textbf {\bibinfo {volume} {61}},\ \bibinfo {pages} {34} (\bibinfo {year} {2024})}\BibitemShut {NoStop}%
\bibitem [{\citenamefont {He}\ \emph {et~al.}(2019{\natexlab{b}})\citenamefont {He}, \citenamefont {Zhang}, \citenamefont {Shams-Ansari}, \citenamefont {Zhu}, \citenamefont {Wang},\ and\ \citenamefont {Marko}}]{HeOL:19}%
  \BibitemOpen
  \bibfield  {author} {\bibinfo {author} {\bibfnamefont {L.}~\bibnamefont {He}}, \bibinfo {author} {\bibfnamefont {M.}~\bibnamefont {Zhang}}, \bibinfo {author} {\bibfnamefont {A.}~\bibnamefont {Shams-Ansari}}, \bibinfo {author} {\bibfnamefont {R.}~\bibnamefont {Zhu}}, \bibinfo {author} {\bibfnamefont {C.}~\bibnamefont {Wang}}, \ and\ \bibinfo {author} {\bibfnamefont {L.}~\bibnamefont {Marko}},\ }\href {\doibase 10.1364/OL.44.002314} {\bibfield  {journal} {\bibinfo  {journal} {Opt. Lett.}\ }\textbf {\bibinfo {volume} {44}},\ \bibinfo {pages} {2314} (\bibinfo {year} {2019}{\natexlab{b}})}\BibitemShut {NoStop}%
\bibitem [{\citenamefont {Harper}\ \emph {et~al.}(1999)\citenamefont {Harper}, \citenamefont {Cabral}, \citenamefont {Andricacos}, \citenamefont {Gignac}, \citenamefont {Noyan}, \citenamefont {Rodbell},\ and\ \citenamefont {Hu}}]{Harper:99}%
  \BibitemOpen
  \bibfield  {author} {\bibinfo {author} {\bibfnamefont {J.~M.~E.}\ \bibnamefont {Harper}}, \bibinfo {author} {\bibfnamefont {J.}~\bibnamefont {Cabral}, \bibfnamefont {C.}}, \bibinfo {author} {\bibfnamefont {P.~C.}\ \bibnamefont {Andricacos}}, \bibinfo {author} {\bibfnamefont {L.}~\bibnamefont {Gignac}}, \bibinfo {author} {\bibfnamefont {I.~C.}\ \bibnamefont {Noyan}}, \bibinfo {author} {\bibfnamefont {K.~P.}\ \bibnamefont {Rodbell}}, \ and\ \bibinfo {author} {\bibfnamefont {C.~K.}\ \bibnamefont {Hu}},\ }\href {\doibase 10.1063/1.371086} {\bibfield  {journal} {\bibinfo  {journal} {Journal of Applied Physics}\ }\textbf {\bibinfo {volume} {86}},\ \bibinfo {pages} {2516} (\bibinfo {year} {1999})},\ \Eprint {http://arxiv.org/abs/https://pubs.aip.org/aip/jap/article-pdf/86/5/2516/19205205/2516\_1\_online.pdf} {https://pubs.aip.org/aip/jap/article-pdf/86/5/2516/19205205/2516\_1\_online.pdf} \BibitemShut {NoStop}%
\bibitem [{\citenamefont {Holzgrafe}\ \emph {et~al.}(2024)\citenamefont {Holzgrafe}, \citenamefont {Puma}, \citenamefont {Cheng}, \citenamefont {Warner}, \citenamefont {Shams-Ansari}, \citenamefont {Shankar},\ and\ \citenamefont {Lon\v{c}ar}}]{Holzgrafe:24}%
  \BibitemOpen
  \bibfield  {author} {\bibinfo {author} {\bibfnamefont {J.}~\bibnamefont {Holzgrafe}}, \bibinfo {author} {\bibfnamefont {E.}~\bibnamefont {Puma}}, \bibinfo {author} {\bibfnamefont {R.}~\bibnamefont {Cheng}}, \bibinfo {author} {\bibfnamefont {H.}~\bibnamefont {Warner}}, \bibinfo {author} {\bibfnamefont {A.}~\bibnamefont {Shams-Ansari}}, \bibinfo {author} {\bibfnamefont {R.}~\bibnamefont {Shankar}}, \ and\ \bibinfo {author} {\bibfnamefont {M.}~\bibnamefont {Lon\v{c}ar}},\ }\href@noop {} {\bibfield  {journal} {\bibinfo  {journal} {Opt. Express}\ }\textbf {\bibinfo {volume} {32}},\ \bibinfo {pages} {3619} (\bibinfo {year} {2024})}\BibitemShut {NoStop}%
\bibitem [{iee(2017)}]{ieee:17}%
  \BibitemOpen
  \href {\doibase 10.1109/IEEESTD.2017.8207825} {\enquote {\bibinfo {title} {{IEEE} standard for ethernet - amendment 10: Media access control parameters, physical layers, and management parameters for 200 gb/s and 400 gb/s operation},}\ }\bibinfo {howpublished} {IEEE Std 802.3bs-201} (\bibinfo {year} {2017})\BibitemShut {NoStop}%
\bibitem [{\citenamefont {Ivanov}\ \emph {et~al.}(2016)\citenamefont {Ivanov}, \citenamefont {Häger}, \citenamefont {Brännström}, \citenamefont {Graell~i Amat}, \citenamefont {Alvarado},\ and\ \citenamefont {Agrell}}]{Ivanov:16}%
  \BibitemOpen
  \bibfield  {author} {\bibinfo {author} {\bibfnamefont {M.}~\bibnamefont {Ivanov}}, \bibinfo {author} {\bibfnamefont {C.}~\bibnamefont {Häger}}, \bibinfo {author} {\bibfnamefont {F.}~\bibnamefont {Brännström}}, \bibinfo {author} {\bibfnamefont {A.}~\bibnamefont {Graell~i Amat}}, \bibinfo {author} {\bibfnamefont {A.}~\bibnamefont {Alvarado}}, \ and\ \bibinfo {author} {\bibfnamefont {E.}~\bibnamefont {Agrell}},\ }\href {\doibase 10.1109/TIT.2016.2543740} {\bibfield  {journal} {\bibinfo  {journal} {IEEE Trans. Inf. Theory}\ }\textbf {\bibinfo {volume} {62}},\ \bibinfo {pages} {3011} (\bibinfo {year} {2016})}\BibitemShut {NoStop}%
\bibitem [{\citenamefont {Hu}\ \emph {et~al.}(2022)\citenamefont {Hu}, \citenamefont {Borkowski}, \citenamefont {Lefevre}, \citenamefont {Cho}, \citenamefont {Buchali}, \citenamefont {Bonk}, \citenamefont {Schuh}, \citenamefont {De~Leo}, \citenamefont {Habegger}, \citenamefont {Destraz}, \citenamefont {Del~Medico}, \citenamefont {Duran}, \citenamefont {Tedaldi}, \citenamefont {Funck}, \citenamefont {Fedoryshyn}, \citenamefont {Leuthold}, \citenamefont {Heni}, \citenamefont {Baeuerle},\ and\ \citenamefont {Hoessbacher}}]{Hu:22}%
  \BibitemOpen
  \bibfield  {author} {\bibinfo {author} {\bibfnamefont {Q.}~\bibnamefont {Hu}}, \bibinfo {author} {\bibfnamefont {R.}~\bibnamefont {Borkowski}}, \bibinfo {author} {\bibfnamefont {Y.}~\bibnamefont {Lefevre}}, \bibinfo {author} {\bibfnamefont {J.}~\bibnamefont {Cho}}, \bibinfo {author} {\bibfnamefont {F.}~\bibnamefont {Buchali}}, \bibinfo {author} {\bibfnamefont {R.}~\bibnamefont {Bonk}}, \bibinfo {author} {\bibfnamefont {K.}~\bibnamefont {Schuh}}, \bibinfo {author} {\bibfnamefont {E.}~\bibnamefont {De~Leo}}, \bibinfo {author} {\bibfnamefont {P.}~\bibnamefont {Habegger}}, \bibinfo {author} {\bibfnamefont {M.}~\bibnamefont {Destraz}}, \bibinfo {author} {\bibfnamefont {N.}~\bibnamefont {Del~Medico}}, \bibinfo {author} {\bibfnamefont {H.}~\bibnamefont {Duran}}, \bibinfo {author} {\bibfnamefont {V.}~\bibnamefont {Tedaldi}}, \bibinfo {author} {\bibfnamefont {C.}~\bibnamefont {Funck}}, \bibinfo {author} {\bibfnamefont {Y.}~\bibnamefont {Fedoryshyn}}, \bibinfo {author} {\bibfnamefont {J.}~\bibnamefont {Leuthold}},
  \bibinfo {author} {\bibfnamefont {W.}~\bibnamefont {Heni}}, \bibinfo {author} {\bibfnamefont {B.}~\bibnamefont {Baeuerle}}, \ and\ \bibinfo {author} {\bibfnamefont {C.}~\bibnamefont {Hoessbacher}},\ }\href {\doibase 10.1109/JLT.2022.3172246} {\bibfield  {journal} {\bibinfo  {journal} {J. Light. Technol.}\ }\textbf {\bibinfo {volume} {40}},\ \bibinfo {pages} {3338} (\bibinfo {year} {2022})}\BibitemShut {NoStop}%
\bibitem [{\citenamefont {Berikaa}\ \emph {et~al.}(2023)\citenamefont {Berikaa}, \citenamefont {Alam}, \citenamefont {Li}, \citenamefont {Bernal}, \citenamefont {Krueger}, \citenamefont {Pittalà},\ and\ \citenamefont {Plant}}]{Berikaa:23}%
  \BibitemOpen
  \bibfield  {author} {\bibinfo {author} {\bibfnamefont {E.}~\bibnamefont {Berikaa}}, \bibinfo {author} {\bibfnamefont {M.~S.}\ \bibnamefont {Alam}}, \bibinfo {author} {\bibfnamefont {W.}~\bibnamefont {Li}}, \bibinfo {author} {\bibfnamefont {S.}~\bibnamefont {Bernal}}, \bibinfo {author} {\bibfnamefont {B.}~\bibnamefont {Krueger}}, \bibinfo {author} {\bibfnamefont {F.}~\bibnamefont {Pittalà}}, \ and\ \bibinfo {author} {\bibfnamefont {D.~V.}\ \bibnamefont {Plant}},\ }\href {\doibase 10.1109/LPT.2023.3285881} {\bibfield  {journal} {\bibinfo  {journal} {IEEE Photonics Technol. Lett.}\ }\textbf {\bibinfo {volume} {35}},\ \bibinfo {pages} {850} (\bibinfo {year} {2023})}\BibitemShut {NoStop}%
\bibitem [{\citenamefont {Zhang}\ \emph {et~al.}(2022)\citenamefont {Zhang}, \citenamefont {Shao}, \citenamefont {Yang}, \citenamefont {Chen}, \citenamefont {Zhang}, \citenamefont {Shum}, \citenamefont {Zhu}, \citenamefont {Chan}, \citenamefont {Lon\v{c}ar},\ and\ \citenamefont {Wang}}]{Zhang:22}%
  \BibitemOpen
  \bibfield  {author} {\bibinfo {author} {\bibfnamefont {Y.}~\bibnamefont {Zhang}}, \bibinfo {author} {\bibfnamefont {L.}~\bibnamefont {Shao}}, \bibinfo {author} {\bibfnamefont {J.}~\bibnamefont {Yang}}, \bibinfo {author} {\bibfnamefont {Z.}~\bibnamefont {Chen}}, \bibinfo {author} {\bibfnamefont {K.}~\bibnamefont {Zhang}}, \bibinfo {author} {\bibfnamefont {K.-M.}\ \bibnamefont {Shum}}, \bibinfo {author} {\bibfnamefont {D.}~\bibnamefont {Zhu}}, \bibinfo {author} {\bibfnamefont {C.~H.}\ \bibnamefont {Chan}}, \bibinfo {author} {\bibfnamefont {M.}~\bibnamefont {Lon\v{c}ar}}, \ and\ \bibinfo {author} {\bibfnamefont {C.}~\bibnamefont {Wang}},\ }\href {\doibase 10.1364/PRJ.468518} {\bibfield  {journal} {\bibinfo  {journal} {Photon. Res.}\ }\textbf {\bibinfo {volume} {10}},\ \bibinfo {pages} {2380} (\bibinfo {year} {2022})}\BibitemShut {NoStop}%
\bibitem [{\citenamefont {Xu}\ \emph {et~al.}(2020)\citenamefont {Xu}, \citenamefont {He}, \citenamefont {Zhang}, \citenamefont {Jian}, \citenamefont {Pan}, \citenamefont {Liu}, \citenamefont {Chen}, \citenamefont {Meng}, \citenamefont {Chen}, \citenamefont {Li}, \citenamefont {Xiao}, \citenamefont {Yu}, \citenamefont {Yu},\ and\ \citenamefont {Cai}}]{Xu:20}%
  \BibitemOpen
  \bibfield  {author} {\bibinfo {author} {\bibfnamefont {M.}~\bibnamefont {Xu}}, \bibinfo {author} {\bibfnamefont {M.}~\bibnamefont {He}}, \bibinfo {author} {\bibfnamefont {H.}~\bibnamefont {Zhang}}, \bibinfo {author} {\bibfnamefont {J.}~\bibnamefont {Jian}}, \bibinfo {author} {\bibfnamefont {Y.}~\bibnamefont {Pan}}, \bibinfo {author} {\bibfnamefont {X.}~\bibnamefont {Liu}}, \bibinfo {author} {\bibfnamefont {L.}~\bibnamefont {Chen}}, \bibinfo {author} {\bibfnamefont {X.}~\bibnamefont {Meng}}, \bibinfo {author} {\bibfnamefont {H.}~\bibnamefont {Chen}}, \bibinfo {author} {\bibfnamefont {Z.}~\bibnamefont {Li}}, \bibinfo {author} {\bibfnamefont {X.}~\bibnamefont {Xiao}}, \bibinfo {author} {\bibfnamefont {S.}~\bibnamefont {Yu}}, \bibinfo {author} {\bibfnamefont {S.}~\bibnamefont {Yu}}, \ and\ \bibinfo {author} {\bibfnamefont {X.}~\bibnamefont {Cai}},\ }\href {\doibase 10.1038/s41467-020-17806-0} {\bibfield  {journal} {\bibinfo  {journal} {Nat. Commun.}\ }\textbf {\bibinfo {volume} {11}},\ \bibinfo {pages} {3911}
  (\bibinfo {year} {2020})}\BibitemShut {NoStop}%
\bibitem [{\citenamefont {Xu}\ \emph {et~al.}(2022)\citenamefont {Xu}, \citenamefont {Zhu}, \citenamefont {Pittal\`{a}}, \citenamefont {Tang}, \citenamefont {He}, \citenamefont {Ng}, \citenamefont {Wang}, \citenamefont {Ruan}, \citenamefont {Tang}, \citenamefont {Kuschnerov}, \citenamefont {Liu}, \citenamefont {Yu}, \citenamefont {Zheng},\ and\ \citenamefont {Cai}}]{Xu:22}%
  \BibitemOpen
  \bibfield  {author} {\bibinfo {author} {\bibfnamefont {M.}~\bibnamefont {Xu}}, \bibinfo {author} {\bibfnamefont {Y.}~\bibnamefont {Zhu}}, \bibinfo {author} {\bibfnamefont {F.}~\bibnamefont {Pittal\`{a}}}, \bibinfo {author} {\bibfnamefont {J.}~\bibnamefont {Tang}}, \bibinfo {author} {\bibfnamefont {M.}~\bibnamefont {He}}, \bibinfo {author} {\bibfnamefont {W.~C.}\ \bibnamefont {Ng}}, \bibinfo {author} {\bibfnamefont {J.}~\bibnamefont {Wang}}, \bibinfo {author} {\bibfnamefont {Z.}~\bibnamefont {Ruan}}, \bibinfo {author} {\bibfnamefont {X.}~\bibnamefont {Tang}}, \bibinfo {author} {\bibfnamefont {M.}~\bibnamefont {Kuschnerov}}, \bibinfo {author} {\bibfnamefont {L.}~\bibnamefont {Liu}}, \bibinfo {author} {\bibfnamefont {S.}~\bibnamefont {Yu}}, \bibinfo {author} {\bibfnamefont {B.}~\bibnamefont {Zheng}}, \ and\ \bibinfo {author} {\bibfnamefont {X.}~\bibnamefont {Cai}},\ }\href {\doibase 10.1364/OPTICA.449691} {\bibfield  {journal} {\bibinfo  {journal} {Optica}\ }\textbf {\bibinfo {volume} {9}},\ \bibinfo {pages}
  {61} (\bibinfo {year} {2022})}\BibitemShut {NoStop}%
\bibitem [{\citenamefont {Curran}\ \emph {et~al.}(2010)\citenamefont {Curran}, \citenamefont {Ndip}, \citenamefont {Werner}, \citenamefont {Ruttkowski}, \citenamefont {Maiwald}, \citenamefont {Wolf}, \citenamefont {Zoellmer}, \citenamefont {Domann}, \citenamefont {Guttovski}, \citenamefont {Gieser},\ and\ \citenamefont {Reichl}}]{Curran:10}%
  \BibitemOpen
  \bibfield  {author} {\bibinfo {author} {\bibfnamefont {B.}~\bibnamefont {Curran}}, \bibinfo {author} {\bibfnamefont {I.}~\bibnamefont {Ndip}}, \bibinfo {author} {\bibfnamefont {C.}~\bibnamefont {Werner}}, \bibinfo {author} {\bibfnamefont {V.}~\bibnamefont {Ruttkowski}}, \bibinfo {author} {\bibfnamefont {M.}~\bibnamefont {Maiwald}}, \bibinfo {author} {\bibfnamefont {H.}~\bibnamefont {Wolf}}, \bibinfo {author} {\bibfnamefont {V.}~\bibnamefont {Zoellmer}}, \bibinfo {author} {\bibfnamefont {G.}~\bibnamefont {Domann}}, \bibinfo {author} {\bibfnamefont {S.}~\bibnamefont {Guttovski}}, \bibinfo {author} {\bibfnamefont {H.}~\bibnamefont {Gieser}}, \ and\ \bibinfo {author} {\bibfnamefont {H.}~\bibnamefont {Reichl}},\ }\href {\doibase 10.1017/S1759078710000450} {\bibfield  {journal} {\bibinfo  {journal} {Int. J. Microw. Wirel. Technol.}\ }\textbf {\bibinfo {volume} {2}},\ \bibinfo {pages} {273–281} (\bibinfo {year} {2010})}\BibitemShut {NoStop}%
\bibitem [{\citenamefont {Xue}\ \emph {et~al.}(2024)\citenamefont {Xue}, \citenamefont {Xu}, \citenamefont {Ding}, \citenamefont {Ye}, \citenamefont {Qiu}, \citenamefont {Li}, \citenamefont {Liu}, \citenamefont {Li}, \citenamefont {Yuan}, \citenamefont {Wang}, \citenamefont {Zheng},\ and\ \citenamefont {Chen}}]{Xue:24}%
  \BibitemOpen
  \bibfield  {author} {\bibinfo {author} {\bibfnamefont {X.}~\bibnamefont {Xue}}, \bibinfo {author} {\bibfnamefont {Y.}~\bibnamefont {Xu}}, \bibinfo {author} {\bibfnamefont {W.}~\bibnamefont {Ding}}, \bibinfo {author} {\bibfnamefont {R.}~\bibnamefont {Ye}}, \bibinfo {author} {\bibfnamefont {J.}~\bibnamefont {Qiu}}, \bibinfo {author} {\bibfnamefont {G.}~\bibnamefont {Li}}, \bibinfo {author} {\bibfnamefont {S.}~\bibnamefont {Liu}}, \bibinfo {author} {\bibfnamefont {H.}~\bibnamefont {Li}}, \bibinfo {author} {\bibfnamefont {L.}~\bibnamefont {Yuan}}, \bibinfo {author} {\bibfnamefont {B.}~\bibnamefont {Wang}}, \bibinfo {author} {\bibfnamefont {Y.}~\bibnamefont {Zheng}}, \ and\ \bibinfo {author} {\bibfnamefont {X.}~\bibnamefont {Chen}},\ }\href {\doibase 10.48550/arXiv.2412.12556} {\enquote {\bibinfo {title} {High-performance thin-film lithium niobate mach-zehnder modulator on thick silica buffering layer},}\ }\bibinfo {howpublished} {Preprint at https://arxiv.org/abs/2412.12556} (\bibinfo {year} {2024})\BibitemShut
  {NoStop}%
\bibitem [{\citenamefont {Wang}\ \emph {et~al.}(2022)\citenamefont {Wang}, \citenamefont {Chen}, \citenamefont {Ruan}, \citenamefont {Gan}, \citenamefont {Huang}, \citenamefont {Zheng}, \citenamefont {Lu}, \citenamefont {Li}, \citenamefont {Guo}, \citenamefont {Chen},\ and\ \citenamefont {Liu}}]{Wang:22}%
  \BibitemOpen
  \bibfield  {author} {\bibinfo {author} {\bibfnamefont {Z.}~\bibnamefont {Wang}}, \bibinfo {author} {\bibfnamefont {G.}~\bibnamefont {Chen}}, \bibinfo {author} {\bibfnamefont {Z.}~\bibnamefont {Ruan}}, \bibinfo {author} {\bibfnamefont {R.}~\bibnamefont {Gan}}, \bibinfo {author} {\bibfnamefont {P.}~\bibnamefont {Huang}}, \bibinfo {author} {\bibfnamefont {Z.}~\bibnamefont {Zheng}}, \bibinfo {author} {\bibfnamefont {L.}~\bibnamefont {Lu}}, \bibinfo {author} {\bibfnamefont {J.}~\bibnamefont {Li}}, \bibinfo {author} {\bibfnamefont {C.}~\bibnamefont {Guo}}, \bibinfo {author} {\bibfnamefont {K.}~\bibnamefont {Chen}}, \ and\ \bibinfo {author} {\bibfnamefont {L.}~\bibnamefont {Liu}},\ }\href {\doibase 10.1021/acsphotonics.2c00263} {\bibfield  {journal} {\bibinfo  {journal} {ACS Photonics}\ }\textbf {\bibinfo {volume} {9}},\ \bibinfo {pages} {2668} (\bibinfo {year} {2022})}\BibitemShut {NoStop}%
\bibitem [{\citenamefont {Daudlin}\ \emph {et~al.}(2025)\citenamefont {Daudlin}, \citenamefont {Rizzo}, \citenamefont {Lee}, \citenamefont {Khilwani}, \citenamefont {Ou}, \citenamefont {Wang}, \citenamefont {Novick}, \citenamefont {Gopal}, \citenamefont {Cullen}, \citenamefont {Parsons}, \citenamefont {Jang}, \citenamefont {Molnar},\ and\ \citenamefont {Bergman}}]{Daudlin:25}%
  \BibitemOpen
  \bibfield  {author} {\bibinfo {author} {\bibfnamefont {S.}~\bibnamefont {Daudlin}}, \bibinfo {author} {\bibfnamefont {A.}~\bibnamefont {Rizzo}}, \bibinfo {author} {\bibfnamefont {S.}~\bibnamefont {Lee}}, \bibinfo {author} {\bibfnamefont {D.}~\bibnamefont {Khilwani}}, \bibinfo {author} {\bibfnamefont {C.}~\bibnamefont {Ou}}, \bibinfo {author} {\bibfnamefont {S.}~\bibnamefont {Wang}}, \bibinfo {author} {\bibfnamefont {A.}~\bibnamefont {Novick}}, \bibinfo {author} {\bibfnamefont {V.}~\bibnamefont {Gopal}}, \bibinfo {author} {\bibfnamefont {M.}~\bibnamefont {Cullen}}, \bibinfo {author} {\bibfnamefont {R.}~\bibnamefont {Parsons}}, \bibinfo {author} {\bibfnamefont {K.}~\bibnamefont {Jang}}, \bibinfo {author} {\bibfnamefont {A.}~\bibnamefont {Molnar}}, \ and\ \bibinfo {author} {\bibfnamefont {K.}~\bibnamefont {Bergman}},\ }\href {\doibase 10.1038/s41566-025-01633-0} {\bibfield  {journal} {\bibinfo  {journal} {Nat. Photonics}\ } (\bibinfo {year} {2025}),\ 10.1038/s41566-025-01633-0}\BibitemShut {NoStop}%
\bibitem [{\citenamefont {Hua}\ \emph {et~al.}(2025)\citenamefont {Hua}, \citenamefont {Divita}, \citenamefont {Yu}, \citenamefont {Peng}, \citenamefont {Roques-Carmes}, \citenamefont {Su}, \citenamefont {Chen}, \citenamefont {Bai}, \citenamefont {Zou}, \citenamefont {Zhu}, \citenamefont {Xu}, \citenamefont {Lu}, \citenamefont {Di}, \citenamefont {Chen}, \citenamefont {Jiang}, \citenamefont {Wang}, \citenamefont {Ou}, \citenamefont {Zhang}, \citenamefont {Chen}, \citenamefont {Zhang}, \citenamefont {Zhu}, \citenamefont {Kuang}, \citenamefont {Wang}, \citenamefont {Meng}, \citenamefont {Steinman},\ and\ \citenamefont {Shen}}]{Hua:25}%
  \BibitemOpen
  \bibfield  {author} {\bibinfo {author} {\bibfnamefont {S.}~\bibnamefont {Hua}}, \bibinfo {author} {\bibfnamefont {E.}~\bibnamefont {Divita}}, \bibinfo {author} {\bibfnamefont {S.}~\bibnamefont {Yu}}, \bibinfo {author} {\bibfnamefont {B.}~\bibnamefont {Peng}}, \bibinfo {author} {\bibfnamefont {C.}~\bibnamefont {Roques-Carmes}}, \bibinfo {author} {\bibfnamefont {Z.}~\bibnamefont {Su}}, \bibinfo {author} {\bibfnamefont {Z.}~\bibnamefont {Chen}}, \bibinfo {author} {\bibfnamefont {Y.}~\bibnamefont {Bai}}, \bibinfo {author} {\bibfnamefont {J.}~\bibnamefont {Zou}}, \bibinfo {author} {\bibfnamefont {Y.}~\bibnamefont {Zhu}}, \bibinfo {author} {\bibfnamefont {Y.}~\bibnamefont {Xu}}, \bibinfo {author} {\bibfnamefont {C.-k.}\ \bibnamefont {Lu}}, \bibinfo {author} {\bibfnamefont {Y.}~\bibnamefont {Di}}, \bibinfo {author} {\bibfnamefont {H.}~\bibnamefont {Chen}}, \bibinfo {author} {\bibfnamefont {L.}~\bibnamefont {Jiang}}, \bibinfo {author} {\bibfnamefont {L.}~\bibnamefont {Wang}}, \bibinfo {author} {\bibfnamefont
  {L.}~\bibnamefont {Ou}}, \bibinfo {author} {\bibfnamefont {C.}~\bibnamefont {Zhang}}, \bibinfo {author} {\bibfnamefont {J.}~\bibnamefont {Chen}}, \bibinfo {author} {\bibfnamefont {W.}~\bibnamefont {Zhang}}, \bibinfo {author} {\bibfnamefont {H.}~\bibnamefont {Zhu}}, \bibinfo {author} {\bibfnamefont {W.}~\bibnamefont {Kuang}}, \bibinfo {author} {\bibfnamefont {L.}~\bibnamefont {Wang}}, \bibinfo {author} {\bibfnamefont {H.}~\bibnamefont {Meng}}, \bibinfo {author} {\bibfnamefont {M.}~\bibnamefont {Steinman}}, \ and\ \bibinfo {author} {\bibfnamefont {Y.}~\bibnamefont {Shen}},\ }\href {\doibase 10.1038/s41586-025-08786-6} {\bibfield  {journal} {\bibinfo  {journal} {Nature}\ }\textbf {\bibinfo {volume} {640}},\ \bibinfo {pages} {361} (\bibinfo {year} {2025})}\BibitemShut {NoStop}%
\bibitem [{\citenamefont {Ahmed}\ \emph {et~al.}(2025)\citenamefont {Ahmed}, \citenamefont {Baghdadi}, \citenamefont {Bernadskiy}, \citenamefont {Bowman}, \citenamefont {Braid}, \citenamefont {Carr}, \citenamefont {Chen}, \citenamefont {Ciccarella}, \citenamefont {Cole}, \citenamefont {Cooke}, \citenamefont {Desai}, \citenamefont {Dorta}, \citenamefont {Elmhurst}, \citenamefont {Gardiner}, \citenamefont {Greenwald}, \citenamefont {Gupta}, \citenamefont {Husbands}, \citenamefont {Jones}, \citenamefont {Kopa}, \citenamefont {Lee}, \citenamefont {Madhavan}, \citenamefont {Mendrela}, \citenamefont {Moore}, \citenamefont {Nair}, \citenamefont {Om}, \citenamefont {Patel}, \citenamefont {Patro}, \citenamefont {Pellowski}, \citenamefont {Radhakrishnani}, \citenamefont {Sane}, \citenamefont {Sarkis}, \citenamefont {Stadolnik}, \citenamefont {Tymchenko}, \citenamefont {Wang}, \citenamefont {Winikka}, \citenamefont {Wleklinski}, \citenamefont {Zelman}, \citenamefont {Ho}, \citenamefont {Jain}, \citenamefont
  {Basumallik}, \citenamefont {Bunandar},\ and\ \citenamefont {Harris}}]{Ahmed:25}%
  \BibitemOpen
  \bibfield  {author} {\bibinfo {author} {\bibfnamefont {S.~R.}\ \bibnamefont {Ahmed}}, \bibinfo {author} {\bibfnamefont {R.}~\bibnamefont {Baghdadi}}, \bibinfo {author} {\bibfnamefont {M.}~\bibnamefont {Bernadskiy}}, \bibinfo {author} {\bibfnamefont {N.}~\bibnamefont {Bowman}}, \bibinfo {author} {\bibfnamefont {R.}~\bibnamefont {Braid}}, \bibinfo {author} {\bibfnamefont {J.}~\bibnamefont {Carr}}, \bibinfo {author} {\bibfnamefont {C.}~\bibnamefont {Chen}}, \bibinfo {author} {\bibfnamefont {P.}~\bibnamefont {Ciccarella}}, \bibinfo {author} {\bibfnamefont {M.}~\bibnamefont {Cole}}, \bibinfo {author} {\bibfnamefont {J.}~\bibnamefont {Cooke}}, \bibinfo {author} {\bibfnamefont {K.}~\bibnamefont {Desai}}, \bibinfo {author} {\bibfnamefont {C.}~\bibnamefont {Dorta}}, \bibinfo {author} {\bibfnamefont {J.}~\bibnamefont {Elmhurst}}, \bibinfo {author} {\bibfnamefont {B.}~\bibnamefont {Gardiner}}, \bibinfo {author} {\bibfnamefont {E.}~\bibnamefont {Greenwald}}, \bibinfo {author} {\bibfnamefont {S.}~\bibnamefont {Gupta}},
  \bibinfo {author} {\bibfnamefont {P.}~\bibnamefont {Husbands}}, \bibinfo {author} {\bibfnamefont {B.}~\bibnamefont {Jones}}, \bibinfo {author} {\bibfnamefont {A.}~\bibnamefont {Kopa}}, \bibinfo {author} {\bibfnamefont {H.~J.}\ \bibnamefont {Lee}}, \bibinfo {author} {\bibfnamefont {A.}~\bibnamefont {Madhavan}}, \bibinfo {author} {\bibfnamefont {A.}~\bibnamefont {Mendrela}}, \bibinfo {author} {\bibfnamefont {N.}~\bibnamefont {Moore}}, \bibinfo {author} {\bibfnamefont {L.}~\bibnamefont {Nair}}, \bibinfo {author} {\bibfnamefont {A.}~\bibnamefont {Om}}, \bibinfo {author} {\bibfnamefont {S.}~\bibnamefont {Patel}}, \bibinfo {author} {\bibfnamefont {R.}~\bibnamefont {Patro}}, \bibinfo {author} {\bibfnamefont {R.}~\bibnamefont {Pellowski}}, \bibinfo {author} {\bibfnamefont {E.}~\bibnamefont {Radhakrishnani}}, \bibinfo {author} {\bibfnamefont {S.}~\bibnamefont {Sane}}, \bibinfo {author} {\bibfnamefont {N.}~\bibnamefont {Sarkis}}, \bibinfo {author} {\bibfnamefont {J.}~\bibnamefont {Stadolnik}}, \bibinfo {author}
  {\bibfnamefont {M.}~\bibnamefont {Tymchenko}}, \bibinfo {author} {\bibfnamefont {G.}~\bibnamefont {Wang}}, \bibinfo {author} {\bibfnamefont {K.}~\bibnamefont {Winikka}}, \bibinfo {author} {\bibfnamefont {A.}~\bibnamefont {Wleklinski}}, \bibinfo {author} {\bibfnamefont {J.}~\bibnamefont {Zelman}}, \bibinfo {author} {\bibfnamefont {R.}~\bibnamefont {Ho}}, \bibinfo {author} {\bibfnamefont {R.}~\bibnamefont {Jain}}, \bibinfo {author} {\bibfnamefont {A.}~\bibnamefont {Basumallik}}, \bibinfo {author} {\bibfnamefont {D.}~\bibnamefont {Bunandar}}, \ and\ \bibinfo {author} {\bibfnamefont {N.~C.}\ \bibnamefont {Harris}},\ }\href {\doibase 10.1038/s41586-025-08854-x} {\bibfield  {journal} {\bibinfo  {journal} {Nature}\ }\textbf {\bibinfo {volume} {640}},\ \bibinfo {pages} {368} (\bibinfo {year} {2025})}\BibitemShut {NoStop}%
\bibitem [{\citenamefont {Edelstein}\ \emph {et~al.}(2001)\citenamefont {Edelstein}, \citenamefont {Uzoh}, \citenamefont {Cabral}, \citenamefont {DeHaven}, \citenamefont {Buchwalter}, \citenamefont {Simon}, \citenamefont {Cooney}, \citenamefont {Malhotra}, \citenamefont {Klaus}, \citenamefont {Rathore}, \citenamefont {Agarwala},\ and\ \citenamefont {Nguyen}}]{Edelstein:01}%
  \BibitemOpen
  \bibfield  {author} {\bibinfo {author} {\bibfnamefont {D.}~\bibnamefont {Edelstein}}, \bibinfo {author} {\bibfnamefont {C.}~\bibnamefont {Uzoh}}, \bibinfo {author} {\bibfnamefont {C.}~\bibnamefont {Cabral}}, \bibinfo {author} {\bibfnamefont {P.}~\bibnamefont {DeHaven}}, \bibinfo {author} {\bibfnamefont {P.}~\bibnamefont {Buchwalter}}, \bibinfo {author} {\bibfnamefont {A.}~\bibnamefont {Simon}}, \bibinfo {author} {\bibfnamefont {E.}~\bibnamefont {Cooney}}, \bibinfo {author} {\bibfnamefont {S.}~\bibnamefont {Malhotra}}, \bibinfo {author} {\bibfnamefont {D.}~\bibnamefont {Klaus}}, \bibinfo {author} {\bibfnamefont {H.}~\bibnamefont {Rathore}}, \bibinfo {author} {\bibfnamefont {B.}~\bibnamefont {Agarwala}}, \ and\ \bibinfo {author} {\bibfnamefont {D.}~\bibnamefont {Nguyen}},\ }in\ \href {\doibase 10.1109/IITC.2001.930001} {\emph {\bibinfo {booktitle} {Proceedings of the IEEE 2001 International Interconnect Technology Conference (Cat. No.01EX461)}}}\ (\bibinfo {year} {2001})\ pp.\ \bibinfo {pages}
  {9--11}\BibitemShut {NoStop}%
\end{thebibliography}%

\vspace{5mm}

\noindent \textbf{Funding} This work was supported by the EU Horizon Europe EIC transition programme under grant agreement No. 101113260 (HDLN), and funded by the Swiss State Secretariat for Education, Research and lnnovation (SERI).

\vspace{1 EM}

\noindent \textbf{Acknowledgements}
The samples were fabricated in the EPFL Center of MicroNano Technology (CMi) and the Institute of Physics (IPHYS) cleanroom. 

\vspace{1 EM}

\noindent\textbf{Competing Interests}
TJK is co-founder of LUXTELLIGENCE SA, offering electro-optical photonic integrated circuits. The other authors declare no competing interests. 
\vspace{1 EM}

\end{document}